\documentclass{aa}
\usepackage[varg]{txfonts}
\usepackage{amsmath}
\allowdisplaybreaks[4]
\usepackage{amssymb}
\usepackage{graphicx}
\usepackage{float}
\usepackage{subfigure}
\usepackage{leftidx}
\usepackage{CJK}
\usepackage{esint}
\usepackage{bm}
\usepackage{color}
\usepackage{tabu}
\usepackage{booktabs}
\usepackage{threeparttable}
\usepackage{multirow}
\usepackage{longtable}
\usepackage{rotating}
\usepackage{makecell}
\usepackage{soul}
\usepackage{ulem}

\usepackage{url}
\usepackage[colorlinks, linkcolor=blue, anchorcolor=blue, citecolor=blue, filecolor=blue, menucolor=blue, runcolor=blue, urlcolor=blue]{hyperref}
\begin{document}
  
  \title{Chemical modeling of the complex organic molecules in the extended region around Sagittarius B2}
  \author{Yao Wang\inst{1,2,3}\and Fujun Du\inst{1,2}\and Dmitry Semenov\inst{3,4}\and Hongchi Wang\inst{1,2}\and Juan Li\inst{5}}
  \institute{Purple Mountain Observatory and Key Laboratory of Radio Astronomy, Chinese Academy of Sciences, 10 Yuanhua Road, 210023 Nanjing, PR China; \email{wangyao@pmo.ac.cn, hcwang@pmo.ac.cn}
  \and
School of Astronomy and Space Science, University of Science and Technology of China, 96 Jinzhai Road, 230026 Hefei, PR China
  \and
Max Planck Institute for Astronomy, K{\"o}nigstuhl 17, 69117 Heidelberg, Germany
\and
Department of Chemistry, Ludwig Maximilian University, Butenandtstr. 5-13, 81377 Munich, Germany
\and
Department of Radio Science and Technology, Shanghai Astronomical Observatory, Chinese Academy of Sciences, 80 Nandan Road, 200030 Shanghai, PR China
  }
  \date{Received date month year / Accepted date month year}
  

  \abstract{The chemical differentiation of seven complex organic molecules (COMs) in the extended region around Sagittarius B2 (Sgr~B2) has been previously observed: CH$_2$OHCHO, CH$_3$OCHO, t-HCOOH, C$_2$H$_5$OH, and CH$_3$NH$_2$ were detected both in the extended region and near the hot cores Sgr~B2(N) and Sgr~B2(M), while CH$_3$OCH$_3$ and C$_2$H$_5$CN were only detected near the hot cores. The density and temperature in the extended region are relatively low in comparison with Sgr~B2(N) and Sgr~B2(M). Different desorption mechanisms, including photodesorption, reactive desorption, and shock heating, and a few other mechanisms have been proposed to explain the observed COMs in the cold regions. However, they fail to explain the deficiency of CH$_3$OCH$_3$ and C$_2$H$_5$CN in the extended region around Sgr~B2. }{Based on known physical properties of the extended region around Sgr~B2, we explored under what physical conditions the chemical simulations can fit the observations and explain the different spatial distribution of these seven species in the extended region around Sgr~B2. }{We used the macroscopic Monte Carlo method to perform a detailed parameter space study. A static physical model and an evolving physical model including a cold phase and a warm-up phase were used, respectively. The fiducial models adopt the observed physical parameters except for the local cosmic ray ionization rate $\zeta_{\mathrm{CR}}$. In addition to photodesorption that is included in all models, we investigated how chain reaction mechanism, shocks, an X-ray burst, enhanced reactive desorption and low diffusion barriers could affect the results of chemical modeling. }{All gas-grain chemical models based on static physics cannot fit the observations, except for the high abundances of CH$_3$NH$_2$ and C$_2$H$_5$CN in some cases. The simulations based on evolving physical conditions can fit six COMs when $T\sim 30-60 \, \mathrm{K}$ in the warm-up phase, but the best-fit temperature is still higher than the observed dust temperature of 20~K. The best agreement between the simulations and all seven observed COMs at a lower temperature $T\sim 27 \, \mathrm{K}$ is achieved by considering a short-duration $\approx 10^2$~yr X-ray burst with $\zeta_{\mathrm{CR}}=1.3 \times 10^{-13} \, \mathrm{s}^{-1}$ at the early stage of the warm-up phase when it still has a temperature of  $20 \, \mathrm{K}$. The reactive desorption is the key mechanism for producing these COMs and inducing the low abundances of CH$_3$OCH$_3$ and C$_2$H$_5$CN.}{We conclude that the evolution of the extended region around Sgr~B2 may have begun with a cold, $T\le10\, \mathrm{K}$ phase followed by a warm-up phase. When its temperature reached about $T\sim20\, \mathrm{K}$, an X-ray flare from the Galactic black hole Sgr~A* with a short duration of no more than 100 years was acquired, affecting strongly the Sgr~B2 chemistry. The observed COMs in Sgr~B2 are able to retain their observed abundances only several hundred years after such a flare, which could imply that such short-term X-rays flares occur relatively often, likely associated with the accretion activity of the Sgr~A* source. }

   
   \maketitle

  \section{Introduction} \label{section1}
  
  \par{}Over 200 species have been detected to date in the interstellar and circumstellar medium \citep{2018ApJS..239...17M}. Those that contain six or more atoms are defined as complex organic molecules (COMs) \citep{2009ARA&A..47..427H}. Among all these species, about one-third were first detected in Sagittarius~B2 (Sgr~B2) \citep{2018ApJS..239...17M}, which remains the best ``hunting ground'' for new COMs \citep{2013A&A...559A..47B}. Sgr~B2 is the most massive star-forming region in our Galaxy. It is located at 8.34 $\pm$ 0.16~kpc from the Sun \citep{2014ApJ...783..130R}, and the projected distance from the Galactic center is 107~pc (or 43.4\arcmin) \citep{2016A&A...588A.143S}. Due to its proximity to the supermassive black hole Sgr~A* at the Galactic center, Sgr~B2 is strongly irradiated by the ultraviolet radiation (UV) field and cosmic ray particles (CRPs). The far-ultraviolet (FUV) radiation field has a strength of $G_0=10^3-10^4$ \citep{2004ApJ...600..214G, 2004A&A...427..217R} (in units of the average local UV interstellar field by Habing with $G_0=1$). The cosmic ray ionization rate is $\zeta_{\mathrm{CR}}=1-11 \times 10^{-14} \, \mathrm{s}^{-1}$ \citep{2015ApJ...800...40I, 2016A&A...585A.105L}, which is two to three orders of magnitude higher than the local CRP ionization rate \citep{1968ApJ...152..971S}. Sgr~B2 consists of three different regions: 1) the high-mass protoclusters Sgr~B2(N) and Sgr~B2(M) with density $n_{\mathrm{H_2}} \sim 10^7\, \mathrm{cm^{-3}}$, 2) a moderate density region extending to about 2.5 pc $\times$ 5.0 pc (or 60.7\arcsec $\times$ 121.3\arcsec) around the hot cores with $n_{\mathrm{H_2}} \sim 10^5\, \mathrm{cm^{-3}}$, and 3) a low-density envelope that extends to about 38~pc (or 15.4\arcmin) with $n_{\mathrm{H_2}} \sim 10^3\, \mathrm{cm^{-3}}$ \citep{1993A&A...276..445H, 1995A&A...294..667H, 2016A&A...588A.143S}. The gas temperature of the extended low-density envelope is about 65~K, and can be as high as several hundred K near Sgr~B2(N) and Sgr~B2(M) \citep{2013A&A...556A.137E}. In contrast, the dust temperature is lower than the gas temperature: the extended envelope, Sgr~B2(N), and Sgr~B2(M) have $T_{\mathrm{dust}}$ of about 20~K, 28~K, and 34~K, respectively \citep{2013A&A...556A.137E}. 
  
  \par{}Molecular species are mostly detected in the hot cores. For example, \citet{2013A&A...559A..47B} have identified 56 species toward Sgr~B2(N) and 46 species toward Sgr~B2(M) with the IRAM 30m telescope. However, emission distributions for some other COMs have been found to be extended over an arcminute scale, covering the region with low density and temperature \citep{2020ARA&A..58..727J}. This has been revealed for glycolaldehyde (CH$_2$OHCHO), propenal (C$_2$H$_3$CHO), propanal (C$_2$H$_5$CHO), cyclopropenone (c-H$_2$C$_3$O), acetamide (CH$_3$CONH$_2$), cyanoformaldehyde (CNCHO), ethanimine (CH$_3$CHNH), and E-cyanomethanimine (HNCHCN) with GBT observations \citep{2004ApJ...613L..45H, 2004ApJ...610L..21H, 2006ApJ...643L..25H, 2006ApJ...642..933H, 2008ApJ...675L..85R, 2013ApJ...765L...9L, 2013ApJ...765L..10Z}, and for methanol (CH$_3$OH), methyl cyanide (CH$_3$CN), methyl acetylene (CH$_3$CCH), formamide (NH$_2$CHO), acetaldehyde (CH$_3$CHO), and cyanodiacetylene (HC$_5$N) with Mopra observations \citep{2008MNRAS.386..117J, 2011MNRAS.411.2293J}. Recently, \citet{2020MNRAS.492..556L} have performed a large-scale mapping of several COMs around Sgr~B2 with the Arizona Radio Observatory (ARO) 12m telescope. Based on the spatial distributions, glycolaldehyde (CH$_2$OHCHO), methyl formate (CH$_3$OCHO), formic acid (t-HCOOH), ethanol (C$_2$H$_5$OH), and methylamine (CH$_3$NH$_2$) can be classified as the ``extended'' molecules since they are distributed over several arcminutes around Sgr~B2(N) and Sgr~B2(M), while dimethyl ether (CH$_3$OCH$_3$) and ethyl cyanide (C$_2$H$_5$CN) can be classified as the ``compact'' molecules since they can only be detected near Sgr~B2(N) and Sgr~B2(M). 
  
  \par{}In addition to the extended region around Sgr~B2, COMs have also been detected in other cold environments in the local ISM, such as pre-stellar cores \citep{2012A&A...541L..12B, 2014ApJ...795L...2V, 2016ApJ...830L...6J} or cold envelopes around low-mass protostars \citep{2010ApJ...716..825O, 2012ApJ...759L..43C, 2014ApJ...791...29J, 2017ApJ...841..120B}. Until recently the presence of COMs in the gas phase at low temperatures ($\lesssim 10-30$~K) has provided a puzzling challenge for astrochemical models. In astrochemistry, the relatively large amount of COMs in hot regions can be understood as follows. During the early stages of star formation, when the temperature is low, many atoms and molecules can be adsorbed on the grain surfaces. Then they can either thermally hop over the surface and react with each other to form more complex molecules via the Langmuir-Hinshelwood mechanism or can react more slowly inside bulk ice via radical chemistry driven by UV photons \citep{2009ARA&A..47..427H}. Next, molecular ices can be desorbed back into the gas phase through either thermal desorption when the dust temperature becomes high enough or via non-thermal desorption processes, including UV-photodesorption, photodesorption driven by cosmic ray particles \citep{2010ApJ...716..825O, 2012ApJ...759L..43C, 2014ApJ...795L...2V}, and reactive desorption driven by exothermic reactions \citep{2007A&A...467.1103G, 2013ApJ...769...34V}. In addition, shocks can strip ice mantles from dust grains and enrich the gas composition with COMs \citep{2006A&A...455..971R, 2008ApJ...672..352R, 2019ApJ...881...32B}. In addition, surface formation of COMs may proceed via the Eley-Rideal and van der Waals complex induced reaction mechanisms \citep{2015MNRAS.447.4004R}. In the so-called chain reaction mechanism, the products of radical-radical reactions on the grain surface can subsequently react with the species that lie right beneath without undergoing thermal diffusion, which can also drive COM chemistry \citep{2016ApJ...819..145C}. A similar mechanism in which the products of photodissociation on the grain surface can immediately react with other species nearby has also been proposed \citep{2019ApJ...884...69G, 2020ApJS..249...26J}. \citet{2018ApJ...861...20S} have suggested that the cosmic-ray bombardment of grain ice mantles can enhance production of COMs from simple ices under cold conditions. More recently, new gas-phase reactions that can produce COMs have been proposed, but they still invoke non-thermal desorption of the COM precursors \citep{2015MNRAS.449L..16B}. 
  
  \par{}Although these mechanisms can explain why many COMs are observed toward Sgr~B2(N) and Sgr~B2(M) and in the cold extended region, it is not fully clear why CH$_3$OCH$_3$ and C$_2$H$_5$CN are deficient in the extended region, as shown by \citet{2020MNRAS.492..556L}. For example, the higher abundance of CH$_3$OCH$_3$ in comparison with CH$_3$OCHO can be explained by different probabilities of reactive desorption upon their formation \citep[$a_{\mathrm{RRK}}$,][]{2007A&A...467.1103G,2013ApJ...769...34V}, by taking new gas-phase routes \citep{2015MNRAS.449L..16B}, by the Eley-Rideal and the van der Waals complex induced reaction mechanisms \citep{2015MNRAS.447.4004R}, or by the chain reaction mechanism \citep{2016ApJ...819..145C}. \citet{2020ApJS..249...26J} have explored the impact of different non-thermal diffusion mechanisms to the formation of CH$_3$OCHO and CH$_3$OCH$_3$ in their rate equation-based simulations. They have found that only models including three-body and excited three-body reactions (similar to the chain reaction mechanism; \citep{2016ApJ...819..145C}) can reproduce higher abundance of CH$_3$OCHO with respect to CH$_3$OCH$_3$, but detailed studies of this kind are lacking for other COMs.
  
  \par{}In this paper we explore in detail under what physical conditions the chemical simulations can fit the distributions of seven COMs (CH$_2$OHCHO, CH$_3$OCHO, t-HCOOH, C$_2$H$_5$OH, CH$_3$NH$_2$, CH$_3$OCH$_3$, and C$_2$H$_5$CN) in Sgr~B2 region. In Sect.~\ref{section2} we describe the chemical models and the physical model grids for simulating the extended region of Sgr~B2. In Sect.~\ref{section3} we present the model results and the best-fit models compared with the observations. In Sect.~\ref{section4} we discuss the reason why the best-fit models with some specific physical parameters can fit the observations, and how it occurs in Sgr~B2. In Sect.~\ref{section5} we present our conclusions.

  \section{Model} \label{section2}
      \subsection{Chemical model} \label{section2.1}
           
      \begin{table*}
          \caption{Main parameters considered in our modeling.}
          \label{table1}
          \begin{center}
          \resizebox{\textwidth}{!}{
          \begin{tabular}{ccccccccccc}\\
          \hline
          \hline
            $n_{\mathrm{H}}$ (cm$^{-3}$)&$A_{\mathrm{V}}$ (mag)&$t_{\mathrm{cold}}$\tablefootmark{a} (yr)&$t_{\mathrm{warmup}}$\tablefootmark{a} (yr)&$T_{\mathrm{max}}$\tablefootmark{a} (K)&$T_{\mathrm{gas}}$\tablefootmark{b} (K)&$T_{\mathrm{dust}}$\tablefootmark{b} (K)&$\chi$\tablefootmark{c}&$G_0$ \tablefootmark{c}&$\zeta_{\mathrm{CR}}$\tablefootmark{d} (s$^{-1}$)&$G_0'$\tablefootmark{d} \\
            \hline
            &&&&&&&1&1&$\bm{1.3\times 10^{-17}}$&$\bm{10^{-4}}$ \\
            $\bm{2\times 10^{3}}$\tablefootmark{e,f}&2&$1\times 10^{5}$&$1\times 10^{3}$&10&10&10&10&10&$1.3\times 10^{-16}$&$10^{-3}$ \\
            $2\times 10^{4}$&5&$2\times 10^{5}$&$1\times 10^{4}$&20&20&$\bm{20}$\tablefootmark{i}&$10^{2}$&$10^{2}$&$1.3\times 10^{-15}$&$10^{-2}$ \\
            $2\times 10^{5}$&$\bm{10}$\tablefootmark{g}&$\bm{3\times 10^{5}}$\tablefootmark{g}&$5\times 10^{4}$&30&30&30&$10^{3}$&$10^{3}$&$1.3\times 10^{-14}$&$10^{-1}$ \\
            $2\times 10^{6}$&25&$4\times 10^{5}$&$1\times 10^{5}$&50&50&&$\bm{10^{4}}$&$\bm{10^{4}}$\tablefootmark{j}&$1.3\times 10^{-13}$\tablefootmark{k}&1 \\
            $2\times 10^{7}$&50&$5\times 10^{5}$&$\bm{2\times 10^{5}}$\tablefootmark{h}&70&$\bm{65}$\tablefootmark{i}&&&&& \\
            &$10^{2}$&$6\times 10^{5}$&$3\times 10^{5}$&100&100&&&&& \\
            &$10^{3}$&$7\times 10^{5}$&$4\times 10^{5}$&150&&&&&& \\
            &&$8\times 10^{5}$&$5\times 10^{5}$&$\bm{200}$\tablefootmark{h}&&&&&& \\
            &&$9\times 10^{5}$&$6\times 10^{5}$&300&&&&&& \\
            &&$1\times 10^{6}$&$7\times 10^{5}$&400&&&&&& \\
            &&&$8\times 10^{5}$&500&&&&&& \\
            &&&$9\times 10^{5}$&700&&&&&& \\
            &&&$1\times 10^{6}$&1000&&&&&& \\
          \hline
          \end{tabular}}
        \end{center}
        \tablefoot{
          \tablefoottext{a}{$t_{\mathrm{cold}}$, $t_{\mathrm{warmup}}$, and $T_{\mathrm{max}}$ are time spans of the cold and warm-up phases and maximum temperature that are used in the evolving physical models. }
          \tablefoottext{b}{$T_{\mathrm{gas}}$ and $T_{\mathrm{dust}}$ are gas and dust temperatures that are used in the static models. }
          \tablefoottext{c}{$\chi$ and $G_0$ are used to calculate the rate coefficients of photodissociation and photodesorption induced by UV photons, respectively. }
          \tablefoottext{d}{$\zeta_{\mathrm{CR}}$ and $G_0'$ are used to calculate the rate coefficients of photodissociation and photodesorption induced by cosmic ray particles, respectively. }
          \tablefoottext{e}{Boldface indicates the fiducial values adopted in our simulations. }
          \tablefoottext{f}{\citet{1993A&A...276..445H, 1995A&A...294..667H, 2016A&A...588A.143S}. }
          \tablefoottext{g}{\citet{2019A&A...622A.185W}.}
          \tablefoottext{h}{\citet{2006A&A...457..927G, 2019A&A...622A.185W}.}
          \tablefoottext{i}{\citet{2013A&A...556A.137E}. }
          \tablefoottext{j}{\citet{2004ApJ...600..214G, 2004A&A...427..217R}. }
          \tablefoottext{k}{\citet{2015ApJ...800...40I, 2016A&A...585A.105L}. }
        }
      \end{table*}

      \begin{table}
        \caption{Initial abundances \citep{2013ApJ...765...60G}. }
        \label{table2}
        \begin{center}
          \begin{tabular}{ll}\\
          \hline
          \hline
            Species&$n_{\mathrm{X}}/n_{\mathrm{H}}$\tablefootmark{a}\\
            \hline
            H$_{2}$&4.99(-1)\\
            H&2.00(-3)\\
            He&9.00(-2)\\
            C&1.40(-4)\\
            N&7.50(-5)\\
            O&3.20(-4)\\
            S&8.00(-8)\\
            Na&2.00(-8)\\
            Mg&7.00(-9)\\
            Si&8.00(-9)\\
            P&3.00(-9)\\
            Cl&4.00(-9)\\
            Fe&3.00(-9)\\
          \hline\\
          \end{tabular}\\
          \tablefoot{\tablefoottext{a}{$a(b)=a\times10^b .$}}
        \end{center}
      \end{table}

\par{}For the extended cold region of Sgr~B2, non-thermal desorption must be very important to release  COMs produced by surface chemistry into the gas phase. Since Sgr~B2 is located nearby the Galactic Center, the local UV radiation field should be strong enough to maintain a high photodesorption efficiency, hence the photodesorption mechanism is considered in all models. According to \citet{2009A&A...496..281O} and \citet{2014ApJ...787..135C}, the UV photodesorption rate coefficients induced by the interstellar radiation field and the cosmic ray particles are respectively  
          \begin{align}
          k_{\mathrm{PD-ISRF}}=G_0 F_0 e^{-\gamma A_\mathrm{V}} \pi r^2 Y_{\mathrm{pd}}  \label{equation1}
          \end{align}     
and
         \begin{align}
          k_{\mathrm{PD-CR}}=G_0' F_0 \pi r^2 Y_{\mathrm{pd}},  \label{equation2}
          \end{align}
where $G_0$ is the external UV radiation scaling factor with a value of 1 for the local ISM, $G_0'$ is the local UV radiation scaling factor induced by the CRPs with a value of $10^{-4}$ \citep{2004A&A...415..203S}, and $F_0$ is the interstellar UV radiation field with a value of $10^8\, \mathrm{photons \, cm^{-2} \, s^{-1}}$. We adopt different values of $G_0$ and $G_0'$ spanning four orders of magnitude. Table \ref{table1} lists these values, and more details are presented in Sect.~\ref{section2.2}. The exponential factor $\gamma$ is a measure of the UV extinction relative to the visual extinction with a value of 2 \citep{1991ApJS...77..287R, 2007ApJ...662L..23O}, and $r$ is the grain radius. The parameter $Y_{\mathrm{pd}}$ is the photodesorption yield, and here it is calculated by Equation~(4), (5) and (6) described in \citet{2009ApJ...693.1209O}. These equations have been derived to describe photodesorption of H$_2$O molecules, since water is the main component on the grain ices; however, we utilize these equations for all kinds of surface molecules in our model. 

\par{}Other mechanisms that can lead to abundant COMs in the gas phase are also included. We adopt the chain reaction mechanism \citep{2016ApJ...819..145C} in some simulations to investigate whether this mechanism can help us to fit the observations. We assume that all species produced by reactions on the grain can participate in this mechanism, not only the products of Langmuir-Hinshelwood reactions, but also the products of photodissociation and accretion. A similar mechanism for the products of photodissociation is included by \citet{2019ApJ...884...69G} in the simulations of ice chemistry in cometary nuclei, and the chain reaction mechanism of the products of accretion is in fact the Eley-Rideal mechanism. The products of Langmuir-Hinshelwood reactions, photodissociation, and absorption should affect the results differently through the chain reaction mechanism. Thus, three types of models are simulated, and each one considers only one particular type of product that can participate in the chain reaction mechanism. The fourth type of our model simulates all three types of products participating in the chain reaction mechanism. We also consider in a rough manner shock action in some chemical models by increasing the temperature within a short timescale characteristic of the shock passage. In addition, we consider an X-ray burst in some models by adopting much larger values of $\zeta_{\mathrm{CR}}$ and $G_0'$ since it can ionize and dissociate species just like a large flux of cosmic ray particles. In addition to all these various processes, the reactive desorption is considered in our models with $a_{\mathrm{RRK}}=0.01$, but we use some larger values (such as 0.1) for testing as well. Furthermore, the diffusion barrier $E_{\mathrm{b}}$ can strongly influence the Langmuir-Hinshelwood reactions. In our models the ratio of the diffusion barrier $E_{\mathrm{b}}$ to the desorption energy $E_{\mathrm{d}}$ is set to be 0.5. Some models are also simulated using a smaller value 0.3. 

\par{}In this work the two-phase gas-grain model is used as the standard chemical kinetics model. We tried to use a three-phase model in our study, as suggested by \citet{1993MNRAS.263..589H} and \citet{2013ApJ...762...86V}, as well as a multi-phase model, as suggested by \citet{2018ApJ...869..165L} and \citet{2019A&A...622A.185W}, but found that these sophisticated models cannot maintain enough COMs in the gas phase in all the considered cases. This is mainly because in multi-phase models COM ices are assumed to be trapped by many monolayers of the bulk water ice, and hence cannot efficiently desorb unless a grain is heated above $\sim 100-150$~K or stripped of ices by shocks. A recent laboratory study by \citet{2020PhRvL.124v1103P} demonstrated that interstellar grains can be very porous and covered by just several monolayers of ice, thus challenging this widespread assumption. In the rest of this paper, we only discuss the results calculated by the two-phase model. 

\par{}We combine the gas-grain chemical reaction network used by \citet{2014Sci...345.1584B} and the network used by \citet{2019A&A...622A.185W} together to cover more COM species and COM reactions. As a result, the combined gas-grain chemical reaction network includes 9710 reactions and 947 species. In addition, we adopt the competition mechanism to calculate the reaction rates for the two surface reactions, CO + H $\rightarrow$ HCO and H$_2$CO + H $\rightarrow$ H$_3$CO \citep{2012ApJ...759..147C, 2019A&A...622A.185W}. The other parameters in the simulations are listed in Table~1 by \citet{2019A&A...622A.185W}. The value of the sticking coefficient is set to be 1 as the standard value, instead of 0.5. The initial abundances that have been used to simulate the evolution of COMs are shown in Table \ref{table2} \citep[e.g.][etc.]{2013ApJ...765...60G, 2014Sci...345.1584B, 2017A&A...601A..49B, 2017A&A...601A..48G, 2019A&A...628A..27B}.
      
\par{}The macroscopic Monte Carlo method used by \citet{2019A&A...622A.185W} is utilized to simulate all these chemical kinetics models. The fractional abundance of a species is defined as the ratio of the population of the species to the total population of H nuclei. In each calculation, the minimum resolvable fractional abundance can only reach $\sim10^{-12}$ \citep{2009ApJ...691.1459V, 2019A&A...622A.185W}. In order to reduce the stochastic fluctuation, we adopt Equation~(8) in \citet{2009ApJ...691.1459V} to calculate the average fractional abundances within chosen time steps. This method can also make the minimum resolvable fractional abundance several orders of magnitude lower than 10$^{-12}$.

\subsection{Physical model} \label{section2.2}
\par{}In the simulations we adopt two types of simple physical models: a static model without any physical evolution, and a homogeneous one-point evolving model including two stages, a cold phase and a warm-up phase \citep{2006A&A...457..927G, 2008ApJ...682..283G, 2008ApJ...681.1385H, 2011ApJ...743..182H, 2013ApJ...765...60G, 2019A&A...622A.185W}. To investigate under what physical conditions the evolution of these seven COMs can fit the observations, we run a grid of models in which we adjust values of several key parameters. These parameters and their values are listed in Table~\ref{table1}, and the fiducial values are in boldface. The fixed gas $T_{\mathrm{gas}}$ and dust $T_{\mathrm{dust}}$ temperatures are used in the static models. The temperature in the cold phase of the evolving models is also fixed. In addition, the evolving models use the timescales of the cold phase $t_{\mathrm{cold}}$ and warm-up phase $t_{\mathrm{warmup}}$ as well as the maximum temperature reached at the end of the warm-up phase $T_{\mathrm{max}}$. We vary high-energy UV radiation field intensities $\chi$ and $G_0$ by the same multiple compared to their smallest values. The same assumption is used to scale the cosmic ray ionization rate $\zeta_{\mathrm{CR}}$ and the corresponding CRP-induced photodissociation and photodesorption rate $G_0'$. Thus, we only discuss the changes of $G_0$ and $\zeta_{\mathrm{CR}}$ in the next sections. The fiducial value of $n_{\mathrm{H}}$, $T_{\mathrm{gas}}$, $T_{\mathrm{dust}}$, and $G_0$ are derived from observations, while the fiducial value of $A_{\mathrm{V}}$, $t_{\mathrm{cold}}$, $t_{\mathrm{warmup}}$, and $T_{\mathrm{max}}$ which can also guarantee to produce enough COMs (see Sect. \ref{section3}), are suggested by \citet{2019A&A...622A.185W}. For $\zeta_{\mathrm{CR}}$, if we adopt a high $\zeta_{\mathrm{CR}}=1-11 \times 10^{-14} \, \mathrm{s}^{-1}$ from the observations, it is difficult to preserve COMs in the gas phase for an extended time because of too strong ionization and dissociation (see Sect. \ref{section3}), and it also makes our Monte Carlo simulations very time-consuming. Thus, a commonly used value $\zeta_{\mathrm{CR}}=1.3 \times 10^{-17} \, \mathrm{s}^{-1}$ is set to be the fiducial value. Hereafter we define fiducial model~1 as the static model with $n_{\mathrm{H}}=2\times10^3 \, \mathrm{cm^{-3}}$, $A_{\mathrm{V}}=10 \, \mathrm{mag}$, $T_{\mathrm{gas}}=65 \, \mathrm{K}$, $T_{\mathrm{dust}}=20 \, \mathrm{K}$, $\chi=10^4$, $G_0=10^4$, $\zeta_{\mathrm{CR}}=1.3\times10^{-17} \, \mathrm{s^{-1}}$, and $G_0'=10^{-4}$; all these parameters remain constant. Similarly, fiducial model~2 is the evolving model with $n_{\mathrm{H}}=2\times10^3 \, \mathrm{cm^{-3}}$, $A_{\mathrm{V}}=10 \, \mathrm{mag}$, $t_{\mathrm{cold}}=3\times10^5 \, \mathrm{yr}$, $t_{\mathrm{warmup}}=2\times10^5 \, \mathrm{yr}$, $T_{\mathrm{max}}=200 \, \mathrm{K}$, $\chi=10^4$, $G_0=10^4$, $\zeta_{\mathrm{CR}}=1.3\times10^{-17} \, \mathrm{s^{-1}}$, and $G_0'=10^{-4}$. In addition, $t_{\mathrm{total}}$ is labeled as the entire evolutionary time span in the simulations. 

\par{}The homogeneous one-point evolving model is described in more detail by \citet{2006A&A...457..927G} (also see \citet{2008ApJ...682..283G, 2008ApJ...681.1385H, 2011ApJ...743..182H, 2013ApJ...765...60G, 2019A&A...622A.185W}), and hence only briefly outlined below. In this evolving model the values of $n_{\mathrm{H}}$, $A_{\mathrm{V}}$, $G_0$, and $\zeta_{\mathrm{CR}}$ are kept constant, and the gas and dust temperatures are assumed to be in equilibrium. In the cold phase the temperature $T=10 \, \mathrm{K}$ and the fiducial value of the timescale $t_{\mathrm{cold}}=3\times 10^5\, \mathrm{yr}$ \citep{2019A&A...622A.185W} are assumed. After the cold phase is over, the temperature during the warm-up phase at a time $t_{\mathrm{cold}}$ + $\Delta t$ is computed as given in \citet{2006A&A...457..927G} and \citet{2019A&A...622A.185W}, 
        \begin{align}
        T=T_0+(T_{\mathrm{max}}-T_0) \left( \frac{\Delta t}{t_{\mathrm{warmup}}} \right) ^n,  \label{equation3}
        \end{align}
        where $T_0=10 \, \mathrm{K}$ is the initial temperature at the beginning of the warm-up phase, $T_{\mathrm{max}}$ is the maximum temperature at the end of the warm-up phase,  $\Delta t$ is the time during the warm-up phase, $t_{\mathrm{warmup}}$ is the timescale of the warm-up phase, and $n=2$ is the order of the heating \citep{2006A&A...457..927G}. Here we set $t_{\mathrm{warmup}}=2\times 10^5\, \mathrm{yr}$ and $T_{\mathrm{max}}=200 \, \mathrm{K}$ as the fiducial values according to \citet{2006A&A...457..927G} and \citet{2019A&A...622A.185W}.

 \section{Results} \label{section3}
  
\par{}In this section the fractional abundances of seven COMs, namely CH$_2$OHCHO, CH$_3$OCHO, t-HCOOH, C$_2$H$_5$OH, CH$_3$NH$_2$, CH$_3$OCH$_3$, and C$_2$H$_5$CN are modeled and discussed. The results of the static models and the evolving models are investigated in Sect. \ref{section3.1} and Sect. \ref{section3.2}, respectively. In Sect. \ref{section3.3} we compare the simulations with the observations to find the best-fit model.

    \subsection{Simulations with the static models} \label{section3.1}

        \begin{figure*}
        \begin{center}
           \includegraphics[width=17.6cm]{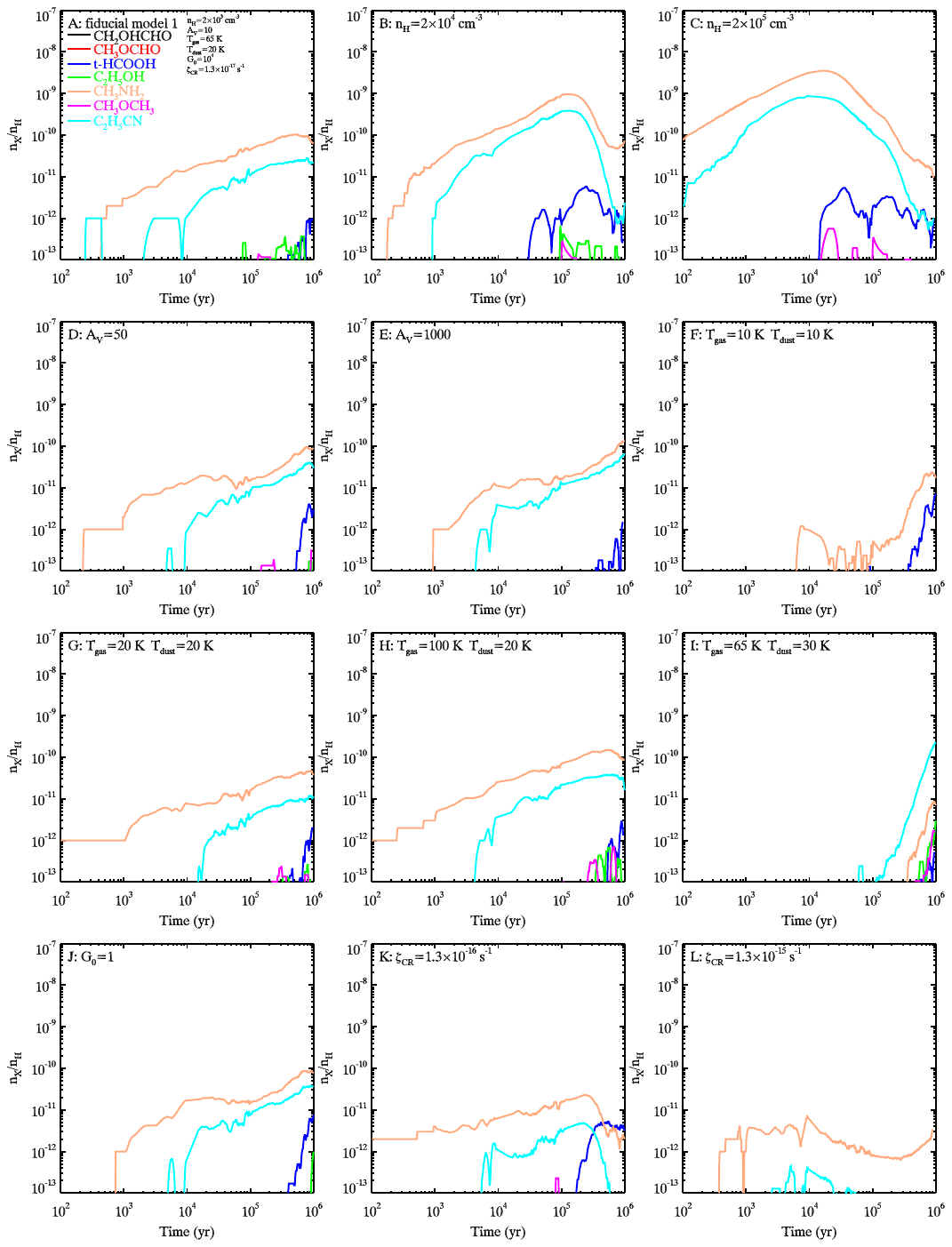}
           \end{center}
        \caption{Time-dependent evolution of the relative gas-phase abundances of CH$_2$OHCHO, CH$_3$OCHO, t-HCOOH, C$_2$H$_5$OH, CH$_3$NH$_2$, CH$_3$OCH$_3$, and C$_2$H$_5$CN in static models. Fiducial model~1 assumes $n_{\mathrm{H}}=2\times10^3 \, \mathrm{cm^{-3}}$, $A_{\mathrm{V}}=10$, $T_{\mathrm{gas}}=65 \, \mathrm{K}$, $T_{\mathrm{dust}}=20 \, \mathrm{K}$, $G_0=10^4$, and $\zeta_{\mathrm{CR}}=1.3\times10^{-17} \, \mathrm{s^{-1}}$. Other static models use different values of each of these parameters at a time. }
        \label{figure1}
        \end{figure*}

        \begin{figure*}
           \begin{center}
           \includegraphics[width=16.6cm]{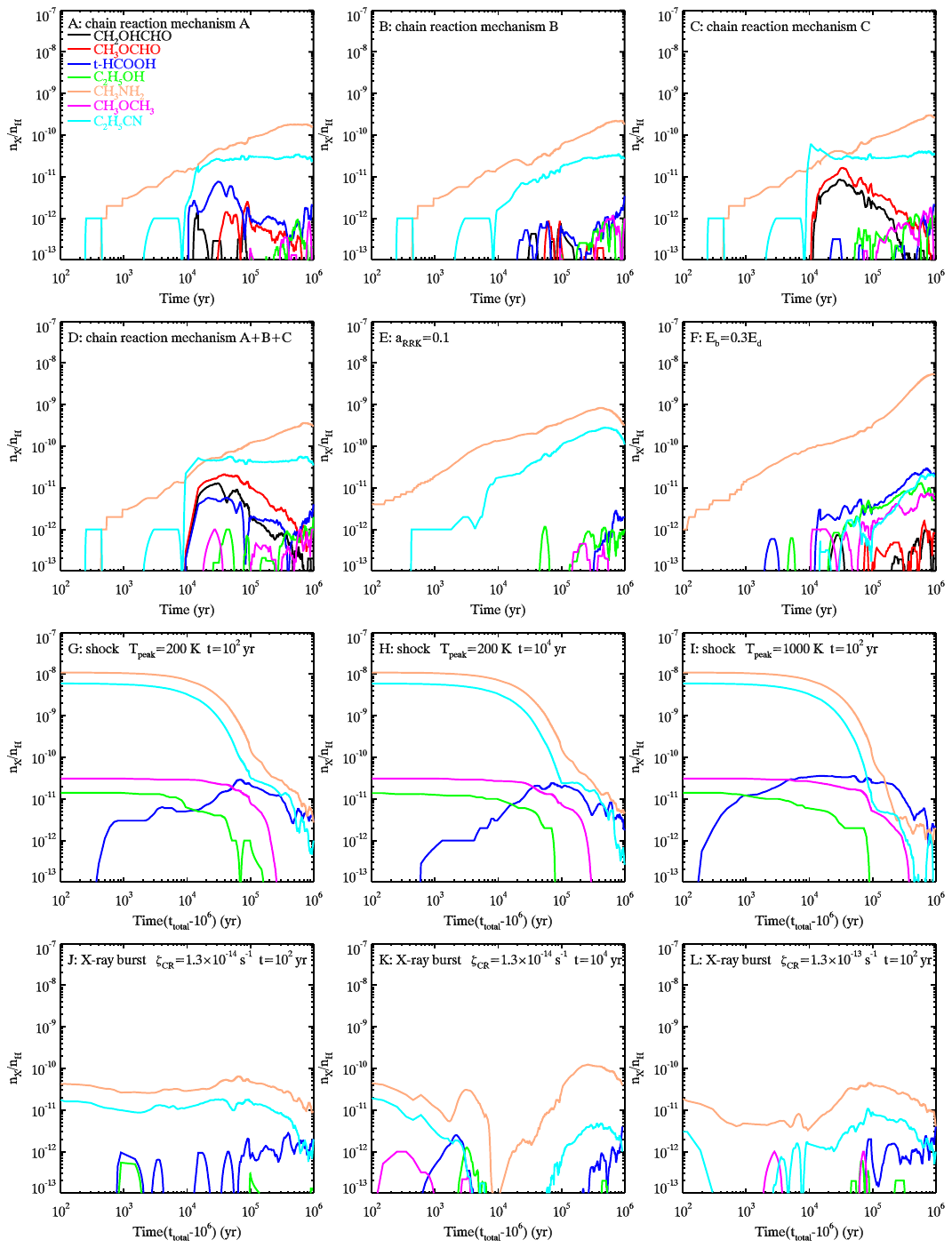}
           \end{center}
           \caption{As in Fig.~\ref{figure1}, but adopting additional mechanisms on the basis of fiducial model~1. In panels~A-C the chain reaction mechanism for the products of Langmuir-Hinshelwood reactions, photodissociation, and adsorption is considered, respectively. In panel~D the chain reaction mechanism for all three types of products is considered. In panel~E $a_{\mathrm{RRK}}=0.1$ is used. In panel~F $E_{\mathrm{b}}=0.3E_{\mathrm{d}}$ is used. In panels~G-L the abscissae start with the abundances at $t_{\mathrm{total}}=10^6\, \mathrm{yr}$ in fiducial model~1, which correspond to $t_{\mathrm{total}}-10^6$ yr. In panels G-I a case with a shock is considered, assuming peak temperatures of $T_{\mathrm{peak}}=200 \, \mathrm{K}$ or 1000~K, and shock duration of $t=10^2\, \mathrm{yr}$ or $10^4\, \mathrm{yr}$. In panels~J-L a case with an X-ray burst is considered, which is described in the models by the elevated  cosmic ray ionization rate $\zeta_{\mathrm{CR}}=1.3\times10^{-14} \, \mathrm{s^{-1}}$ or $1.3\times10^{-13} \, \mathrm{s^{-1}}$ during a short timescale of  $t=10^2\, \mathrm{yr}$ or $10^4\, \mathrm{yr}$, respectively. After the shock or the X-ray burst, all parameters are assumed to be the same as in fiducial model~1. }
           \label{figure2}
        \end{figure*}  
  
  \par{}First of all, we verified whether gas-phase relative abundances of CH$_2$OHCHO, CH$_3$OCHO, t-HCOOH, C$_2$H$_5$OH, and CH$_3$NH$_2$ can become higher than CH$_3$OCH$_3$ and C$_2$H$_5$CN in the static models. As outlined in Table~\ref{table1}, we changed one key parameter for each static model, starting from fiducial model~1. Figure~\ref{figure1} shows the time-dependent evolution of the abundances of these seven COMs in the basic cases of the static models. In addition, Figure~\ref{figure2} depicts the time-dependent evolution of these COMs in the more complex static models considering the chain reaction mechanism, the shock passage, the X-ray burst, a larger value of $a_{\mathrm{RRK}}$, or a smaller value of $E_{\mathrm{b}}$ on the basis of fiducial model~1.

  \subsubsection{Impact of various key parameters on the outcome of static chemical models}\label{section3.1.1}
  
  \par{}According to the median abundances from observations, the relative gas-phase abundances of CH$_3$OCH$_3$ and C$_2$H$_5$CN are lower than 10$^{-12}$, while the abundances of the other five COMs are higher than 10$^{-11}$ (see Sect. \ref{section3.3}). In fiducial model~1 ($n_{\mathrm{H}}=2\times10^3 \, \mathrm{cm^{-3}}$, $A_{\mathrm{V}}=10$, $T_{\mathrm{gas}}=65 \, \mathrm{K}$, $T_{\mathrm{dust}}=20 \, \mathrm{K}$, $G_0=10^4$, and $\zeta_{\mathrm{CR}}=1.3\times10^{-17} \, \mathrm{s^{-1}}$; see panel A in Fig. \ref{figure1}) only the abundances of CH$_3$NH$_2$ and C$_2$H$_5$CN become higher than 10$^{-11}$, while for t-HCOOH it can only reach 10$^{-12}$, and for the other four COMs they are lower than 10$^{-12}$. Therefore, only CH$_3$NH$_2$ and CH$_3$OCH$_3$ can match the observations while the other five COMs cannot, and these results cannot fit the observations as a whole. It can also be seen that, similar to fiducial model~1, all the other static models fail to fit the observations.

  \par{}The static model with a higher density of $n_{\mathrm{H}}=2\times10^4 \, \mathrm{cm^{-3}}$ (panel B in Fig. \ref{figure1}) results in higher peak abundances of the considered COMs, but the differences disappear after 10$^6$ years. It indicates that larger values of $n_{\mathrm{H}}$ cannot make deficient gaseous COMs become abundant. This is also true for another static model with a higher density of $n_{\mathrm{H}}=2\times10^5 \, \mathrm{cm^{-3}}$ (panel C in Fig. \ref{figure1}); the gas-phase chemistry is mostly affected, while surface processes are not affected effectively. 
  \par{}For the models using extinction factors $A_{\mathrm{V}}=10$, 50, and 1000 (panels A, D, and E in Fig. \ref{figure1}), the differences remain small, since the photodissociation rate remains similarly low when $A_{\mathrm{V}}\ge10$. For the more UV-irradiated cases with $A_{\mathrm{V}}=2$ or 5, the abundances of all seven COMs become lower than 10$^{-13}$, which means COMs are being quickly photodissociated (these two plots are not shown). Thus, $A_{\mathrm{V}}=10$ is a reasonable value for preventing gaseous COMs from being photodissociated. 
  \par{}For the static model with $T_{\mathrm{gas}}=10 \, \mathrm{K}$ and $T_{\mathrm{dust}}=10 \, \mathrm{K}$ (panel F in Fig. \ref{figure1}), only abundant gas-phase CH$_3$NH$_2$ can be produced most of the time. When a higher gas temperature of $T_{\mathrm{gas}}=20 \, \mathrm{K}$ or 100~K is utilized with a higher dust temperature $T_{\mathrm{dust}}=20 \, \mathrm{K}$ (panels G and H in Fig. \ref{figure1}), the results remain similar to fiducial model~1. Similarly, for the static model with $T_{\mathrm{gas}}=65 \, \mathrm{K}$ and $T_{\mathrm{dust}}=30 \, \mathrm{K}$ (panel I in Fig. \ref{figure1}), all considered COMs cannot be efficiently produced except for CH$_3$NH$_2$, whose abundance can reach 10$^{-11}$, and C$_2$H$_5$CN , whose abundance can reach 10$^{-10}$ at $t_{\mathrm{total}}\sim 10^6\, \mathrm{yr}$. These four models indicate that dust temperature $T_{\mathrm{dust}}$ can affect the COM chemistry more strongly, whereas $T_{\mathrm{gas}}$ cannot, meaning that only surface reactions play a key role in producing COMs, not gas-phase reactions.
  \par{}The results from the static model using $G_0=1$ (panel J in Fig. \ref{figure1}) are similar to those from fiducial model~1 using $G_0=10^4$ because the corresponding physical conditions are assumed to be ``dark,'' with $A_{\mathrm{V}}\ge10$. In the cases with the higher cosmic ray ionization rates of $\zeta_{\mathrm{CR}}=1.3\times10^{-16} \, \mathrm{s^{-1}}$ and $1.3\times10^{-15} \, \mathrm{s^{-1}}$ (panels K and L in Fig. \ref{figure1}), the higher the value of $\zeta_{\mathrm{CR}}$, the lower the abundances of these seven COMs. For the cases when even higher cosmic ray ionization rates of $\zeta_{\mathrm{CR}}=1.3\times10^{-14} \, \mathrm{s^{-1}}$ and $1.3\times10^{-13} \, \mathrm{s^{-1}}$ are considered, all seven COM abundances are lower than 10$^{-12}$ due to too enhanced ionization and dissociation (these two plots are not shown). 
  
  \par{}In summary, the abundances of COMs in the static models are mainly influenced by $A_{\mathrm{V}}$, $T_{\mathrm{dust}}$, and $\zeta_{\mathrm{CR}}$. The strength of the local ambient FUV radiation field $G_0$ can influence the results only if visual extinction $A_{\mathrm{V}}$ is low. The variations in gas density $n_{\mathrm{H}}$ and temperature $T_{\mathrm{gas}}$ also do not strongly affect the outcome of the static models.

  \subsubsection{Impact of more exotic mechanisms on the outcome of static chemical models}\label{section3.1.2}
  
  \par{}Since the results of the static models with photodesorption and reactive desorption cannot fit the observations, we consider several additional models that adopt the chain reaction mechanism, a shock passage, an X-ray burst, a larger value of $a_{\mathrm{RRK}}$, or a smaller value of $E_{\mathrm{b}}$ (on the basis of fiducial model~1). Figure \ref{figure2} depicts the time-dependent evolution of the relative abundances in these models. 
  \par{}The static models that consider the chain reaction mechanism for the products of Langmuir-Hinshelwood reactions, photodissociation, and adsorption are respectively labeled chain reaction mechanism A, B, and C (panels A, B, and C in Fig. \ref{figure2}). The static model that considers that all products of these three types of processes can participate in the chain reaction mechanism is labeled chain reaction mechanism A+B+C (panel D in Fig. \ref{figure2}). All seven COM abundances are a little higher than fiducial model~1. However, only the abundance of t-HCOOH in the case of chain reaction mechanism A, the abundances of CH$_2$OHCHO and CH$_3$OCHO in the case of chain reaction mechanism C, and the abundances of CH$_2$OHCHO, CH$_3$OCHO, and t-HCOOH in the case of chain reaction mechanism A+B+C can reach detectable levels of $\sim 10^{-11}$, while the abundance of C$_2$H$_5$OH remains lower than 10$^{-12}$ in all four models. Thus, the addition of the chain reaction mechanism cannot solve the problem. 
  \par{}For the static model where the probability of reactive desorption $a_{\mathrm{RRK}}=0.1$ (panel E in Fig. \ref{figure2}) is assumed to be ten times higher than the fiducial value 0.01, the enhanced reactive desorption increases the abundances of t-HCOOH, C$_2$H$_5$OH, CH$_3$NH$_2$, CH$_3$OCH$_3$, and C$_2$H$_5$CN. However, only the abundances of CH$_3$NH$_2$ and C$_2$H$_5$CN become higher than 10$^{-11}$, while the abundances of the other COMs are still lower than 10$^{-12}$. 
  \par{}For the static model where lower diffusion barriers with $E_{\mathrm{b}}=0.3E_{\mathrm{d}}$ (panel F in Fig. \ref{figure2}) are assumed, the efficiency of Langmuir-Hinshelwood reactions is enhanced. In this case the abundance of C$_2$H$_5$CN decreases to 10$^{-11}$, while the abundances of the other six COMs increase. The abundances of t-HCOOH, C$_2$H$_5$OH, and CH$_3$OCH$_3$ become about 10$^{-11}$, while for CH$_2$OHCHO and CH$_3$OCHO they are nearly 10$^{-12}$. 
  
  \par{}Next, we consider a static model with a shock that occurs at $t_{\mathrm{total}}=10^6\, \mathrm{yr}$ (keeping the rest of the parameters in fiducial model~1). We designate the peak temperature that can be reached during the shock as $T_{\mathrm{peak}}$, and the timescale of the shock passage as $t$. For simplicity, the gas and dust temperature both reach the same $T_{\mathrm{peak}}$ of 200~K or 1000~K during the shock, and the shock lasts for $10^2\, \mathrm{yr}$ or $10^4\, \mathrm{yr}$ (panels G-I in Fig. \ref{figure2}). The onset of the shock is set to be the initial time 0 on the abscissae in panels~G-I in Fig.~\ref{figure2}. The results of these models are similar, which indicates that the exact value of the temperature in the shock between 200~K and 1000 K is not important, and that the timescale of the shock passage between $10^2\, \mathrm{yr}$ and $10^4\, \mathrm{yr}$ is also not that important. The main reason is that the higher temperatures above 200~K during the shock passage lead to almost instantaneous evaporation of almost all ices (at much shorter timescale than $10^2\, \mathrm{yr}$), while simultaneously increasing efficiency of gas-phase reactions (important mainly for simple molecules). The resulting abundances of t-HCOOH, C$_2$H$_5$OH, and CH$_3$OCH$_3$ can reach  $\sim 10^{-11}$, while the abundances of CH$_3$NH$_2$ and C$_2$H$_5$CN can become as high as 10$^{-8}$ for about $10^4\, \mathrm{yr}$ after the shock passage. After that, COM abundances decrease due to freeze out and become lower than 10$^{-11}$. However, even in this case the abundances of CH$_2$OHCHO and CH$_3$OCHO remain too low.
  \par{}Finally, we consider a case with an X-ray burst that can be simulated in our modeling by enhancing the cosmic ray ionization rate to higher values of $1.3 \times 10^{-14} \, \mathrm{s}^{-1}$ or $1.3 \times 10^{-13} \, \mathrm{s}^{-1}$. We again assume that such a burst occurs at $t_{\mathrm{total}}=10^6\, \mathrm{yr}$ in fiducial model~1 and lasts for $10^2\, \mathrm{yr}$ or $10^4\, \mathrm{yr}$. The corresponding results are shown in panels~J-L in Fig.~\ref{figure2}. The relative gas-phase abundances of CH$_3$NH$_2$ and C$_2$H$_5$CN become lower compared to those in fiducial model~1 due to enhanced ionization and dissociation, and the abundances of the other five COMs are still lower than 10$^{-12}$. 
    
  \par{}We conclude that the Monte Carlo chemical simulations based on static physical conditions cannot reliably fit the observations. The results are also hardly influenced when considering more exotic scenarios with a shock passage or an X-ray burst since they are mainly determined by the chemical evolution prior to these temporal phenomena.

      \subsection{Simulations with the evolving models} \label{section3.2}

        \begin{figure*}
        \begin{center}
           \includegraphics[width=16.8cm]{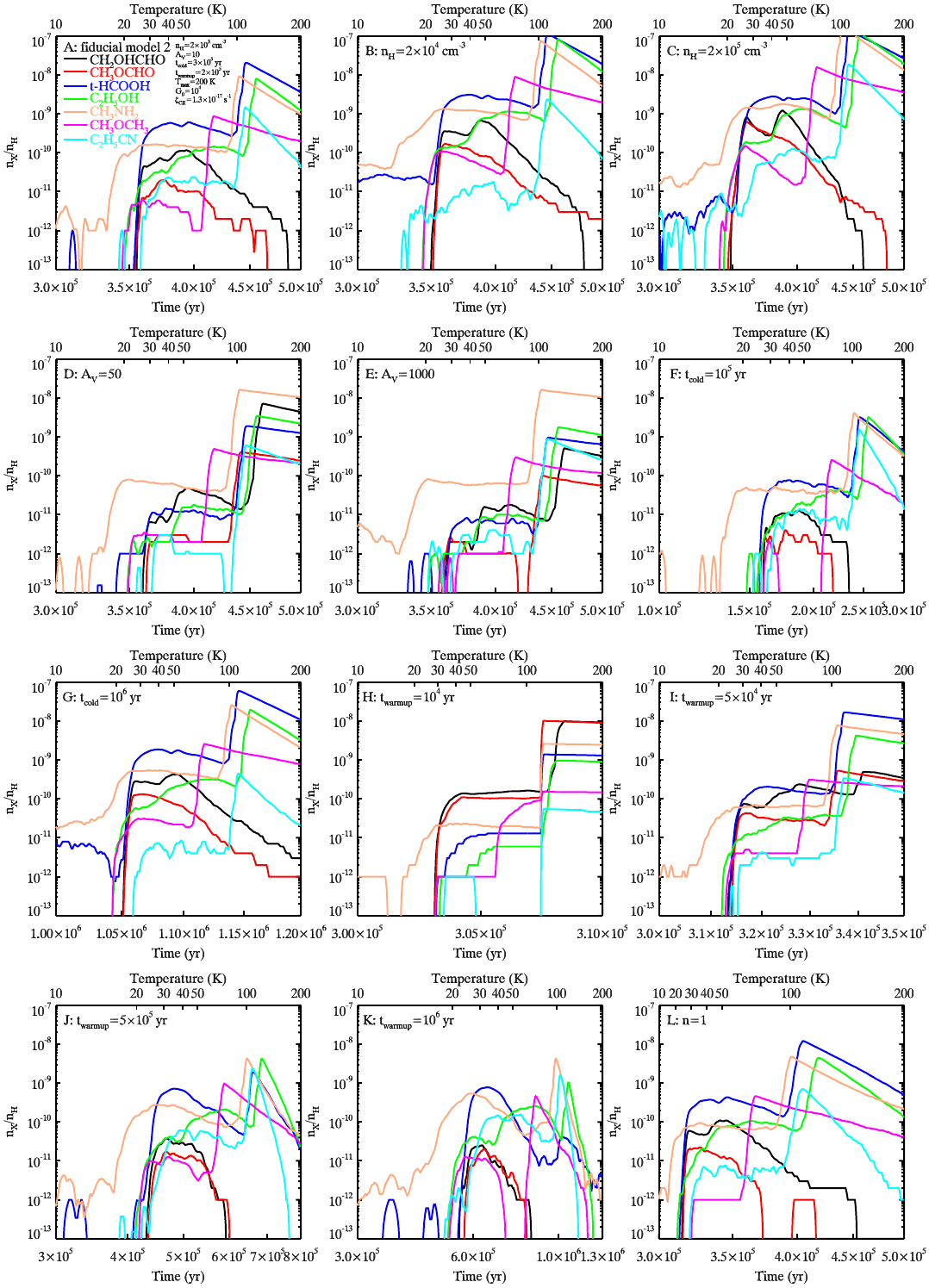}
           \end{center}
            \caption{Time-dependent evolution of the relative gas-phase abundances of CH$_2$OHCHO, CH$_3$OCHO, t-HCOOH, C$_2$H$_5$OH, CH$_3$NH$_2$, CH$_3$OCH$_3$, and C$_2$H$_5$CN in the warm-up phase in evolving models. Fiducial model~2 adopts $n_{\mathrm{H}}=2\times10^3 \, \mathrm{cm^{-3}}$, $A_{\mathrm{V}}=10$, $t_{\mathrm{cold}}=3\times10^5 \, \mathrm{yr}$, $t_{\mathrm{warmup}}=2\times10^5 \, \mathrm{yr}$, $T_{\mathrm{max}}=200 \, \mathrm{K}$, $G_0=10^4$, $\zeta_{\mathrm{CR}}=1.3\times10^{-17} \, \mathrm{s^{-1}}$. Other evolving models adopt a different value of each of these parameters at a time. In panel~L, a linear rather than a quadratic warm-up rate is used in Equation~\ref{equation3} ($n=1$). }
            \label{figure3}
        \end{figure*}
        \addtocounter{figure}{-1}
        \begin{figure*}
        \begin{center}
           \includegraphics[width=16.8cm]{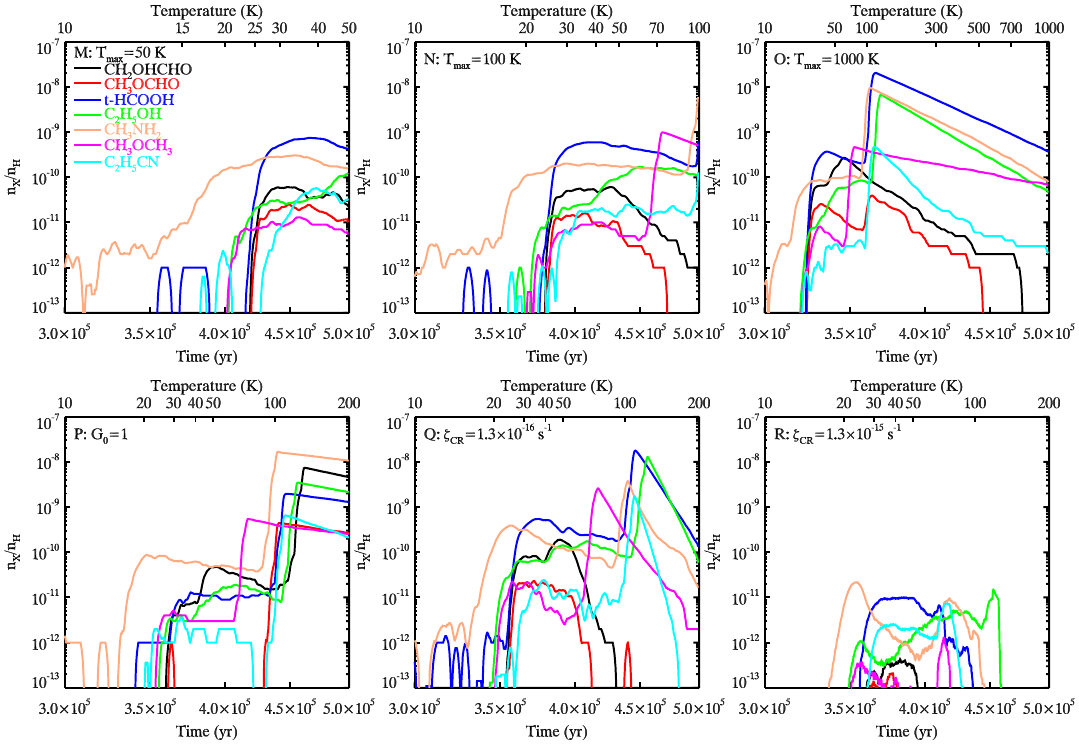}
           \end{center}
            \caption{Continued. }
            \label{figure3_2}
        \end{figure*}
        
        \begin{figure*}
        \begin{center}
           \includegraphics[width=16.2cm]{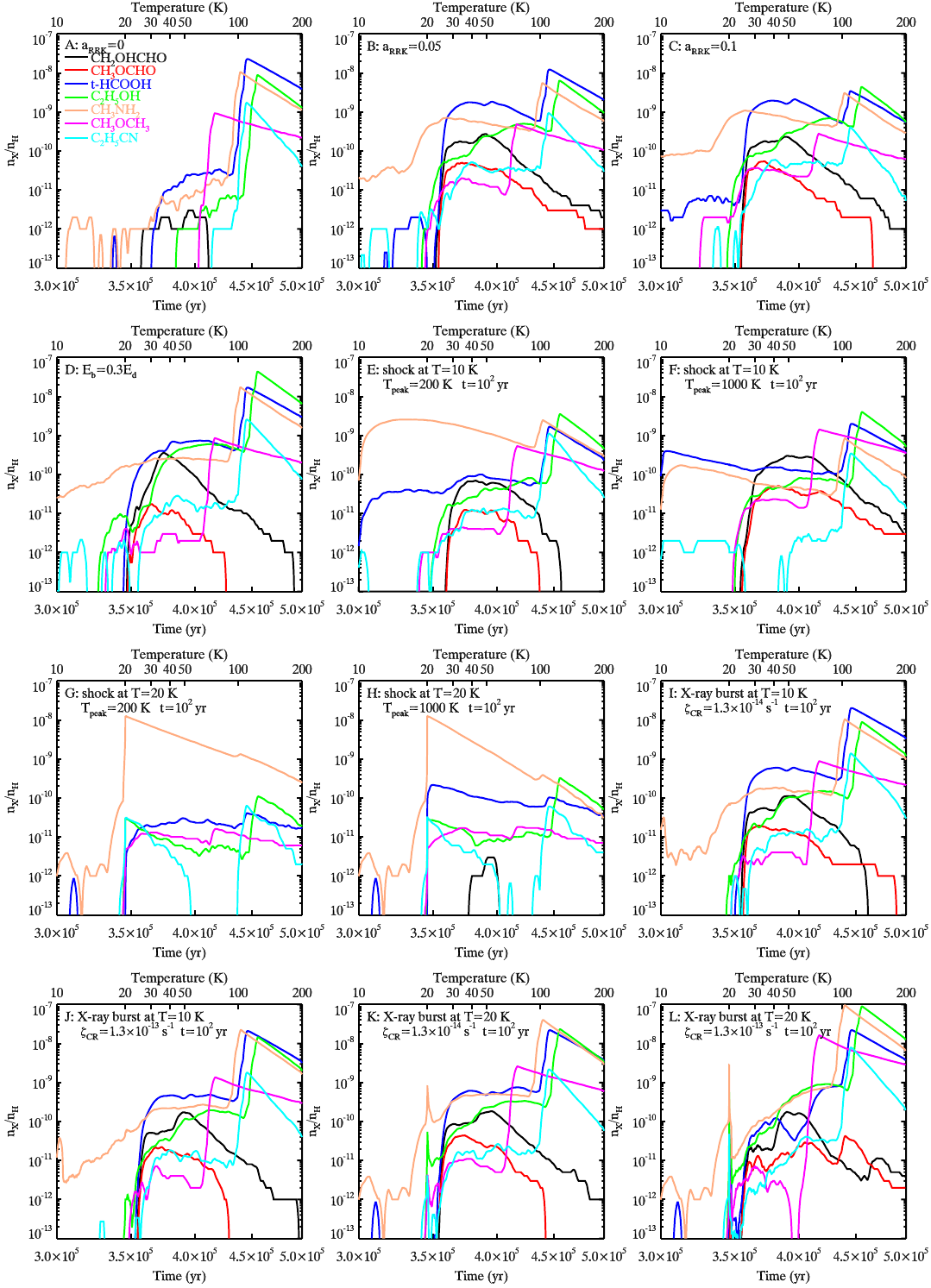}
           \end{center}
            \caption{As in Fig.~\ref{figure3}, but adopting additional mechanisms on the basis of fiducial model~2. In panels~A-C three cases with $a_{\mathrm{RRK}}=0$, 0.05, and 0.1 are considered, respectively. In panel~D a case with faster diffusion rates $E_{\mathrm{b}}=0.3E_{\mathrm{d}}$ is considered. In panels~E-H a scenario with a shock passage is considered, where $T$ = 10~K and 20~K represent the temperature when the shocks occurs in the warm-up phase ($t_{\mathrm{total}}=3\times10^5 \, \mathrm{yr}$ and $3.459\times10^5 \, \mathrm{yr}$, respectively). Peak temperatures $T_{\mathrm{peak}}=200 \, \mathrm{K}$ and 1000 K are the temperatures during shock passage, and $t=10^2\, \mathrm{yr}$ is the duration the passage. In panels~I-L a scenario with an X-ray burst is considered, which is described in the models by the elevated cosmic ray ionization rate of $\zeta_{\mathrm{CR}}=1.3\times10^{-14} \, \mathrm{s^{-1}}$ and $1.3\times10^{-13} \, \mathrm{s^{-1}}$, $T$ = 10~K and 20~K, and the X-ray burst duration is $t=10^2\, \mathrm{yr}$. After the shock passage and the X-ray burst, all parameters are the same as in fiducial model~2. }
            \label{figure4}
        \end{figure*}
        
        \par{}Since the static models failed to fit the observations, we ran a number of evolving chemical models by including a cold phase and a warm-up phase to simulate the abundances of the seven COMs observed in Sgr~B2. 
        
        \subsubsection{Impact of various key parameters on the outcome of evolving chemical models}\label{section3.2.1}
        
        \par{}Similar to the static models, we changed one parameter each time starting from fiducial model~2 according to Table~\ref{table1}, and discuss the results. Figure~\ref{figure3} shows the time-dependent evolution of the relative gas-phase abundances of these seven COMs in evolving models. Only the warm-up phase is shown because in the cold phase the evolution is similar to the static model using $T_{\mathrm{gas}}=10 \, \mathrm{K}$ and $T_{\mathrm{dust}}=10 \, \mathrm{K}$. Overall, the evolution trends in over one-half of the  calculated models are similar. 
        
        \par{}The results of fiducial model~2 ($n_{\mathrm{H}}=2\times10^3 \, \mathrm{cm^{-3}}$, $A_{\mathrm{V}}=10$, $t_{\mathrm{cold}}=3\times10^5 \, \mathrm{yr}$, $t_{\mathrm{warmup}}=2\times10^5 \, \mathrm{yr}$, $T_{\mathrm{max}}=200 \, \mathrm{K}$, $G_0=10^4$, $\zeta_{\mathrm{CR}}=1.3\times10^{-17} \, \mathrm{s^{-1}}$) are depicted in panel~A in Fig. \ref{figure3}. At the start of the warm-up phase, all seven COM abundances are still very low. The abundances of CH$_2$OHCHO, CH$_3$OCHO, t-HCOOH, and C$_2$H$_5$CN start to increase rapidly at $T\sim 25 \, \mathrm{K}$; the abundances of C$_2$H$_5$OH and CH$_3$OCH$_3$ start to increase at $T\sim 20 \, \mathrm{K}$; and the CH$_3$NH$_2$ abundance starts to increase at $T\sim 16 \, \mathrm{K}$. All seven COM abundances remain relatively stable when $T<60 \, \mathrm{K}$. After that the abundances of CH$_2$OHCHO and CH$_3$OCHO decrease, the abundance of CH$_3$OCH$_3$ increases again to reach the peak abundance at $T\sim 60 \, \mathrm{K}$ and then decreases, while for t-HCOOH, C$_2$H$_5$OH, CH$_3$NH$_2$, and C$_2$H$_5$CN the abundances increase again to reach their peak values at different $T\sim 100 \, \mathrm{K}$ (and then they decrease). The peak abundances are caused by the evaporation of the COMs from the grain surfaces. In essence, the abundances of CH$_2$OHCHO, CH$_3$OCHO, t-HCOOH, C$_2$H$_5$OH, and CH$_3$NH$_2$ are higher than the abundances of CH$_3$OCH$_3$ and C$_2$H$_5$CN when the temperature reaches $\sim 30-40 \, \mathrm{K}$, and then they fit the observations reasonably well. When temperatures are higher, $T\sim 40-60 \, \mathrm{K}$, the abundance of C$_2$H$_5$CN is about two times higher than that of CH$_3$OCHO, but the overall results can still fit the observations (see Sect.~\ref{section3.3} below). However, the derived best-fit temperature is a little higher ($\sim$ 37 K or 57 K, see Sect.~\ref{section3.3}) in comparison with the observed dust temperature of 20~K in the extended region of Sgr~B2.      
       
       \par{}When different gas densities are considered, namely $n_{\mathrm{H}}=2\times10^4 \, \mathrm{cm^{-3}}$ and $2\times10^5 \, \mathrm{cm^{-3}}$ (panels B and C in Fig. \ref{figure3}), all seven COM abundances increase and the t-HCOOH, CH$_3$NH$_2$, and CH$_3$OCH$_3$ abundances become higher in comparison with the observations. 
       \par{}For the darker models using $A_{\mathrm{V}}=50$ and 1000 (panels D and E in Fig. \ref{figure3}), seven COM abundances become a little lower when $T\sim 30-60 \, \mathrm{K}$ than those in fiducial model~2. After that, at higher temperatures when these molecules are evaporated from grains, the abundances of t-HCOOH, C$_2$H$_5$OH, CH$_3$OCH$_3$, and C$_2$H$_5$CN decrease, but the abundances of CH$_2$OHCHO, CH$_3$OCHO, and CH$_3$NH$_2$ increase compared to fiducial model~2. In more FUV irradiated cases when $A_{\mathrm{V}}=2$ or 5, all seven COM abundances remain lower than 10$^{-13}$due to a too high intensity of photodissociation (and hence these two plots are not presented). It seems that $A_{\mathrm{V}}=10$ is a reasonable value, as we found for the static models. 
       \par{}The model using a shorter duration of the cold phase $t_{\mathrm{cold}}=10^5 \, \mathrm{yr}$ (panel F in Fig. \ref{figure3}) has lower COM abundances in the warm-up phase, while the model with a longer duration of the cold phase $t_{\mathrm{cold}}=10^6 \, \mathrm{yr}$ (panel G in Fig. \ref{figure3}) has higher COM abundances in the warm-up phase, except for C$_2$H$_5$CN (compared to those in fiducial model~2). It indicates that a longer timescale of the cold phase $t_{\mathrm{cold}}$ can help to produce more key precursors of these COMs before the onset of the warm-up phase. However, the overproduction of t-HCOOH, CH$_3$NH$_2$, and CH$_3$OCH$_3$ in the longer cold-phase model with $t_{\mathrm{cold}}=10^6 \, \mathrm{yr}$ results in a worse fit to the observations (see Sect.~\ref{section3.3}). The COM abundances in the models with other values of the cold-phase duration lie between these two extremes, and the discrepancies are insignificant compared to the results of fiducial model~2. Thus, we conclude that a moderately long cold-phase duration $t_{\mathrm{cold}}\sim 3\times10^5 \, \mathrm{yr}$ is necessary for attaining a good fit to the Sgr~B2 observations.   
       \par{}Next, we adopt four different values for the duration of the warm-up phase $t_{\mathrm{warmup}}$. The results for the model with $t_{\mathrm{warmup}}=10^4 \, \mathrm{yr}$ (panel H in Fig. \ref{figure3}) are very different from those in fiducial model~2. The abundances of CH$_2$OHCHO and CH$_3$OCHO become higher than the abundances of the other five COMs, and both increase when $T\sim 100 \, \mathrm{K}$ because of the evaporation. Similarly, in the model with the warm-up duration of $5\times10^4 \, \mathrm{yr}$ (panel I in Fig. \ref{figure3}), the CH$_2$OHCHO and CH$_3$OCHO abundances continue to increase at higher temperatures compared to their values at $T\sim 30-60 \, \mathrm{K}$ due to more efficient thermal desorption (all the way up to $\sim 100$~K). In contrast, abundances of these two species decrease at $T\sim 50 \, \mathrm{K}$ in the models with longer duration of the warm-up phase, $5\times10^5 \, \mathrm{yr}$ and $10^6 \, \mathrm{yr}$ (panels J and K in Fig. \ref{figure3}). Due to the short timescale of $10^4 \, \mathrm{yr}$, the CH$_2$OHCHO and CH$_3$OCHO molecules produced on the grains are not destroyed by OH radicals when $T\sim 50 \, \mathrm{K}$ (see Sect. \ref{section4.1}) before their evaporation, thus the evaporated CH$_2$OHCHO and CH$_3$OCHO molecules can increase their abundances again when $T\sim 100 \, \mathrm{K}$. In the model using a longer timescale of $5\times10^4 \, \mathrm{yr}$, the longer duration at $T\sim 50 \, \mathrm{K}$ already induces more CH$_2$OHCHO and CH$_3$OCHO molecules on the grains to be destroyed by OH radicals, so their abundances in the gas-phase increase only a little when $T\sim 100 \, \mathrm{K}$. For even longer timescales of $2\times10^5 \, \mathrm{yr}$, $5\times10^5 \, \mathrm{yr}$, and $10^6 \, \mathrm{yr}$, the duration at $T\sim 50 \, \mathrm{K}$ is long enough to destroy most CH$_2$OHCHO and CH$_3$OCHO molecules on the grains, and these molecules in the gas-phase are destroyed by C$^+$ (see Sect. \ref{section4.1}), so the decreased curves occur when $T > 50 \, \mathrm{K}$. The abundances of other five COMs also decrease more strongly at higher temperatures ($T>100 \, \mathrm{K}$) in these two models using $t_{\mathrm{warmup}}=5\times10^5 \, \mathrm{yr}$ and $10^6 \, \mathrm{yr}$ compared to fiducial model~2. The longer duration of the warm-up phase, while allowing COMs to be desorbed into the gas phase, also leads to more efficient destruction of these COMs through the ion-molecule reactions (see Sect.~\ref{section4.1}). Hence $t_{\mathrm{warmup}}$ should be longer than $5\times10^4\, \mathrm{yr}$ and should not be too long, since long timescales of $t_{\mathrm{warmup}}$ can induce more ion-molecule reactions to destroy these COMs in the gas-phase, although it is not obvious in the models using $t_{\mathrm{warmup}}=5\times10^5 \, \mathrm{yr}$ and $10^6 \, \mathrm{yr}$. A moderate value of around $2\times10^5 \, \mathrm{yr}$ is a good choice, and it is better not to deviate from this value too much.
        \par{}In panel~L in Fig. \ref{figure3}, the results for the model with linear slope of the warm-up power law are shown. They are similar to fiducial model~2 with a steeper quadratic warm-up power law if we set the same temperature intervals on the abscissa. This implies that the temperature values are more significant for the chemistry than the pace of the warm-up process. 
        \par{}Next, we consider three models with different maximum temperature values $T_{\mathrm{max}}$ of 50~K, 100~K, and 1000~K (panels M, N, and O in Fig. \ref{figure3_2}). Similarly to the case with a linear warm-up rate, if we set the same temperature intervals on the abscissa the evolution curves in these model look similar to those in fiducial model~2. Hence, the maximum temperature $T_{\mathrm{max}}$ at the end of the warm-up phase is not that important if it is higher than $50-100$~K. 
        \par{}The model with a lower FUV radiation field of $G_0=1$ (panel P in Fig. \ref{figure3_2}) leads to lower COM abundances at $T\sim 30-60 \, \mathrm{K}$; higher peak abundances of CH$_2$OHCHO, CH$_3$OCHO, and CH$_3$NH$_2$; and lower peak abundances of t-HCOOH, C$_2$H$_5$OH, CH$_3$OCH$_3$, and C$_2$H$_5$CN compared to fiducial model 2, which contradicts the observations. For the models using $G_0=10$, 100, and 1000, their results are similar to the model using $G_0=1$ (these plots are not shown). Thus, the high intensity of the UV-photodissociation and photodesorption are needed, with $G_0 \approx 10^4$. 
        \par{}In the evolving model with a higher cosmic ray ionization rate $\zeta_{\mathrm{CR}}=1.3\times10^{-16} \, \mathrm{s^{-1}}$ (panel Q in Fig. \ref{figure3_2}), the results at $T\sim 30-60 \, \mathrm{K}$ are similar to those in fiducial model~2. At later times (and higher temperatures) all seven COM abundances decrease more rapidly due to enhanced ionization and dissociation. In the model with even higher $\zeta_{\mathrm{CR}}=1.3\times10^{-15} \, \mathrm{s^{-1}}$ (panel R in Fig. \ref{figure3_2}), all COM abundances are lower than 10$^{-11}$, and for the extreme ionization models with $\zeta_{\mathrm{CR}}=1.3\times10^{-14} \, \mathrm{s^{-1}}$ and $1.3\times10^{-13} \, \mathrm{s^{-1}}$, the COM abundances are lower than 10$^{-13}$ (these two plots are not shown). Thus, the cosmic ray ionization rate $\zeta_{\mathrm{CR}}$ should not be very high on long timescales, at most several times higher than $1.3\times10^{-16} \, \mathrm{s^{-1}}$. 
        
        \par{}To conclude, $A_{\mathrm{V}}$ and $\zeta_{\mathrm{CR}}$ can affect the results dramatically, $n_{\mathrm{H}}$ and $G_0$ can influence the results to some extent, while $T_{\mathrm{max}}$ is not that important. Timescales of the cold and warm-up phase, $t_{\mathrm{cold}}$ and $t_{\mathrm{warmup}}$, are also not influential if moderate values are selected, such as $t_{\mathrm{cold}}=3\times 10^5\, \mathrm{yr}$ and $t_{\mathrm{warmup}}=2\times 10^5\, \mathrm{yr}$.

        \subsubsection{Impact of more exotic mechanisms on the outcome of evolving chemical models}\label{section3.2.2}
        
        \par{}Although in some evolving models the abundances of CH$_2$OHCHO, CH$_3$OCHO, t-HCOOH, C$_2$H$_5$OH, and CH$_3$NH$_2$ are higher than the abundances of CH$_3$OCH$_3$ and C$_2$H$_5$CN when $T\sim 30-60 \, \mathrm{K}$, all seven COM abundances cannot fit the observed values at the same time moment (see Sect.~\ref{section3.3}). Moreover, the observed dust temperature is also lower than the dust temperature in the evolving models with COM abundances closest to the observed values. To try to achieve a better, more physically correct best fit with a dust temperature of 20~K, we also consider several additional evolving models by considering more exotic mechanisms and parameter values, namely larger values of $a_{\mathrm{RRK}}$, a smaller value of $E_{\mathrm{b}}$, a shock passage, or an X-ray burst (based on fiducial model~2). The time-dependent evolution of the relative COM gas-phase abundances in these models are shown in Fig.~\ref{figure4}. 
                
        \par{}In the model without reactive desorption ($a_{\mathrm{RRK}}=0$, panel A in Fig. \ref{figure4}), all seven COM abundances are low and only the relative abundances of t-HCOOH and CH$_3$NH$_2$ can reach 10$^{-12}$ at $T\sim 20-60 \, \mathrm{K}$. This indicates the significance of the reactive desorption for producing enough COMs in the gas phase when thermal evaporation does not work. In the models with $a_{\mathrm{RRK}}=0.05$ and 0.1 (panels B and C in Fig. \ref{figure4}), the differences in the COM abundances with those in fiducial model 2 are subtle. Thus, the efficiency of the reactive desorption with a reasonable value $a_{\mathrm{RRK}}=0.01$ is required to fit the observations.
        
        \par{}In the model with a lower diffusion barrier $E_{\mathrm{b}}=0.3E_{\mathrm{d}}$ (panel D in Fig. \ref{figure4}), the abundances of CH$_2$OHCHO and CH$_3$NH$_2$ increase, while the abundances of t-HCOOH and CH$_3$OCH$_3$ decrease slightly at $T\sim 30-60 \, \mathrm{K}$ and produce a better fit to the observations than fiducial model~2 (see Sect.~\ref{section3.3}). Thus, the ratio of $E_{\mathrm{b}}$ to $E_{\mathrm{d}}$ is an important parameter, and is likely $\sim 0.3-0.5$, as derived in laboratory experiments \citep{2017SSRv..212....1C}.
        
        \par{}Next, we consider an evolving model with a shock (based on fiducial model~2). The shock passage is assumed to start when $T=10 \, \mathrm{K}$ or 20~K (at $t_{\mathrm{total}}=3\times10^5 \, \mathrm{yr}$ or $3.459\times10^5 \, \mathrm{yr}$) in the warm-up phase, the peak temperature during the shock passage is $T_{\mathrm{peak}}=200 \, \mathrm{K}$ or 1000 K, and the shock passage lasts $t=10^2 \, \mathrm{yr}$ (panels E-H in Fig. \ref{figure4}). After the shock all parameters return to the same values as in fiducial model~2. In the model with the shock happening at $T=10 \, \mathrm{K}$ and with $T_{\mathrm{peak}}=200 \, \mathrm{K}$ (panel E in Fig. \ref{figure4}), the abundance of t-HCOOH becomes higher when $T<25 \, \mathrm{K}$ and then lower when $T>25 \, \mathrm{K}$, but CH$_3$NH$_2$ becomes higher when $T< 100 \, \mathrm{K}$ and then lower when $T>100 \, \mathrm{K}$ in comparison with fiducial model~2. The abundances of the other five COMs change insignificantly. In the case when the shock happens at $T=10 \, \mathrm{K}$ but with higher $T_{\mathrm{peak}}=1000 \, \mathrm{K}$ (panel F in Fig. \ref{figure4}), the abundances of CH$_2$OHCHO, CH$_3$OCHO, and CH$_3$OCH$_3$ become higher, but the abundances of t-HCOOH, CH$_3$NH$_2$, and C$_2$H$_5$CN become lower compared to the results of fiducial model~2 when $T>30 \, \mathrm{K}$. In the models where the shock passage occurs later at $T=20 \, \mathrm{K}$ (panels G and H in Fig. \ref{figure4}) CH$_3$NH$_2$ is overproduced, while the abundances of CH$_2$OHCHO and CH$_3$OCHO are lower than 10$^{-12}$. The longer timescales of the shock ($t=10^3\, \mathrm{yr}$ or $10^4\, \mathrm{yr}$) are also simulated, and their results are similar to the short duration shock passage model using $t=10^2\, \mathrm{yr}$ (these plots are not shown). Therefore, the models with an early shock passage when $T=10 \, \mathrm{K}$ can fit the COM observations in Sgr~B2 at a later moment, when $T\sim 30-60 \, \mathrm{K}$. However, these models cannot reproduce the observations at a more feasible temperature $T=20 \, \mathrm{K}$ (see Sect. \ref{section3.3}).
        
        \par{}Finally, we consider evolving models with an X-ray burst instead of a shock. The outburst also begins early, when $T=10 \, \mathrm{K}$ or 20 K, and leads to temporarily high ionization rates of $\zeta_{\mathrm{CR}}=1.3\times10^{-14} \, \mathrm{s^{-1}}$ or $1.3\times10^{-13} \, \mathrm{s^{-1}}$ (panels I-L in Fig. \ref{figure4}). The duration of the X-ray burst is short,  $t=10^2 \, \mathrm{yr}$. The differences between computed COM abundances of these X-ray burst models with the burst onset at $T=10 \, \mathrm{K}$ (panels I and J in Fig. \ref{figure4}) and the results of fiducial model~2 are small. In contrast, in the case when such an X-ray burst occurs later, at $T=20 \, \mathrm{K}$ (panels K and L in Fig. \ref{figure4}), more C$_2$H$_5$OH but less CH$_3$NH$_2$ and CH$_3$OCH$_3$ are produced at $T=20-30 \, \mathrm{K}$ compared with the outcome of fiducial model~2. This makes the modeling results fit the COM observations in Sgr~B2 at more realistic dust temperatures $T<30 \, \mathrm{K}$ (see Sect.~\ref{section3.3}). The longer timescales of the X-ray burst lead to worse fits due to excessive COM destruction (these plots are not shown).

     \subsection{Comparison with observations} \label{section3.3}
  
   \par{}The ``mean confidence level'' method \citep{2007A&A...467.1103G,2008ApJ...681.1385H, 2019A&A...622A.185W} was adopted to compare the simulations of several evolving models with the observations to quantify the best-fit value of each model. In this method, for species $j$ the confidence level $\kappa_j$ is defined as the agreement between the abundance $X_j$ in the simulation and the observation $X_{\mathrm{obs,}j}$, which is calculated by
        \begin{align}
        \kappa_j=\mathrm{erfc} \left( \frac{|\lg X_j - \lg X_{\mathrm{obs,}j}|}{\sqrt{2}\sigma} \right),  \label{equation4}
        \end{align}
        where erfc is the complementary error function and $\sigma=1$ is the standard deviation. Thus, the value of $\kappa_j$ lies between 0 and 1, and the larger the $\kappa_j$ value, the better agreement between the simulations and the observations. To quantify the global agreement, the average $\kappa$ for all seven COMs is calculated. Furthermore, the maximum of the average $\kappa$ can be obtained; the specific time, the temperature, and the number of fits (which is the number of species that can fit the observations) can also be obtained.

        \begin{sidewaystable*}
        \caption{Observed and simulated fractional abundances of the seven COMs in the extended region of Sgr~B2. }
        \label{table3}
        \centering
          \begin{tabular}{lccccccccccc}\\
          \hline
          \hline

            Model/Observation&CH$_2$OHCHO&CH$_3$OCHO&t-HCOOH&C$_2$H$_5$OH&CH$_3$NH$_2$&CH$_3$OCH$_3$&C$_2$H$_5$CN&Number of fits\tablefootmark{f}&$\kappa$\tablefootmark{g}&$t_{\mathrm{total}}$\tablefootmark{h} (yr)&$T$\tablefootmark{i} (K)\\
            \hline
            \multirow{3}{*}{\makecell{Observation}\tablefootmark{a}}&2.1(-09)\tablefootmark{c}&1.3(-10)&1.0(-10)&1.7(-09)&4.3(-10)&&&&&&\\
            &|&|&|&|&|&<4.2(-12)&<3.0(-11)&-&-&-&-\\
            &1.4(-11)&1.0(-12)&1.3(-12)&2.0(-11)&2.8(-12)&&&&&&\\
            \hline
            Median value\tablefootmark{b}&1.6(-10)&1.9(-11)&1.4(-11)&2.2(-10)&4.6(-11)&4.2(-13)&8.3(-13)&7&-&-&-\\
            Fiducial model 2&6.5(-11)&1.9(-11)&$\textbf{5.3(-10)}$\tablefootmark{d}&3.3(-11)&1.9(-10)&4.0(-12)&1.8(-11)&6&0.473&3.752(05)&36.8\\
            &9.3(-11)&1.0(-11)&$\textbf{5.6(-10)}$&1.1(-10)&1.4(-10)&1.0(-12)&2.3(-11)&6&0.565&3.992(05)&57.2\\
            $n_{\mathrm{H}}=2\times10^4 \, \mathrm{cm^{-3}}$&4.3(-11)&1.8(-11)&$\textbf{1.6(-10)}$&1.1(-10)&$\textbf{1.2(-09)}$&$\textbf{1.0(-10)}$&3.0(-12)&4&0.481&3.547(05)&24.2\\
            $n_{\mathrm{H}}=2\times10^5 \, \mathrm{cm^{-3}}$&4.8(-11)&3.6(-11)&2.2(-11)&8.8(-11)&$\textbf{1.1(-09)}$&$\textbf{7.0(-11)}$&2.0(-12)&5&0.545&3.539(05)&23.8\\
            $A_{\mathrm{V}}=50$&4.6(-11)&3.0(-12)&1.3(-11)&$\textbf{1.6(-11)}$&4.7(-11)&2.0(-12)&1.0(-12)&6&0.667&3.937(05)&51.7\\
            $A_{\mathrm{V}}=1000$&1.7(-11)&1.0(-12)&7.0(-12)&$\textbf{9.0(-12)}$&6.1(-11)&1.0(-12)&2.0(-12)&6&0.539&3.913(05)&49.6\\
            $t_{\mathrm{cold}}=10^5 \, \mathrm{yr}$&$\textbf{6.0(-12)}$&2.0(-12)&4.9(-11)&$\textbf{5.0(-12)}$&4.6(-11)&1.0(-12)&1.0(-12)&5&0.545&1.664(05)&30.9\\
            $t_{\mathrm{cold}}=\times10^6 \, \mathrm{yr}$&2.7(-10)&1.2(-10)&$\textbf{8.9(-10)}$&5.8(-11)&$\textbf{5.2(-10)}$&$\textbf{2.9(-11)}$&1.0(-12)&4&0.450&1.061(06)&27.6\\
            $t_{\mathrm{warmup}}=10^4 \, \mathrm{yr}$&1.3(-10)&1.1(-10)&1.1(-11)&$\textbf{2.0(-12)}$&2.2(-11)&1.0(-12)&7.1(-13)&6&0.678&3.047(05)&52.8\\
            $t_{\mathrm{warmup}}=5\times10^4 \, \mathrm{yr}$&1.6(-10)&2.7(-11)&$\textbf{1.7(-10)}$&3.2(-11)&6.5(-11)&4.0(-12)&2.0(-12)&6&0.637&3.258(05)&60.7\\
            $t_{\mathrm{warmup}}=5\times10^5 \, \mathrm{yr}$&2.3(-11)&1.3(-11)&$\textbf{3.7(-10)}$&1.2(-10)&1.5(-10)&4.0(-12)&3.0(-11)&6&0.466&5.216(05)&47.3\\
            $t_{\mathrm{warmup}}=10^6 \, \mathrm{yr}$&$\textbf{1.0(-12)}$&3.0(-12)&$\textbf{1.1(-10)}$&1.9(-10)&6.9(-11)&6.7(-13)&$\textbf{1.5(-10)}$&4&0.499&7.742(05)&52.7\\
            $n=1$&7.1(-11)&1.8(-11)&$\textbf{4.9(-10)}$&7.3(-11)&1.3(-10)&3.0(-12)&1.1(-11)&6&0.647&3.380(05)&46.1\\
            $T_{\mathrm{max}}=50 \, \mathrm{K}$&3.5(-11)&1.4(-11)&$\textbf{2.8(-10)}$&2.7(-11)&2.7(-10)&$\textbf{8.0(-12)}$&1.0(-12)&5&0.505&4.303(05)&27.0\\
            $T_{\mathrm{max}}=100 \, \mathrm{K}$&2.6(-11)&1.0(-11)&$\textbf{1.4(-10)}$&$\textbf{1.5(-11)}$&1.8(-10)&4.0(-12)&9.4(-13)&5&0.517&3.851(05)&26.3\\
            $T_{\mathrm{max}}=1000 \, \mathrm{K}$&2.0(-10)&1.5(-11)&$\textbf{2.9(-10)}$&5.9(-11)&1.0(-10)&3.0(-12)&2.0(-12)&6&0.632&3.433(05)&56.3\\
            $G_0=1$&4.8(-11)&$\textbf{0}$\tablefootmark{e}&1.1(-11)&$\textbf{1.4(-11)}$&4.9(-11)&3.0(-12)&1.0(-12)&5&0.579&3.940(05)&52.0\\
            $\zeta_{\mathrm{CR}}=1.3\times10^{-16} \, \mathrm{s^{-1}}$&1.8(-10)&1.6(-11)&$\textbf{2.7(-10)}$&1.4(-10)&1.2(-10)&2.0(-12)&1.0(-11)&6&0.629&3.920(05)&50.2\\
            $\zeta_{\mathrm{CR}}=1.3\times10^{-15} \, \mathrm{s^{-1}}$&$\textbf{7.1(-14)}$&$\textbf{7.5(-13)}$&5.3(-12)&$\textbf{9.4(-14)}$&4.2(-12)&4.1(-13)&1.1(-12)&4&0.434&3.650(05)&30.0\\
            $a_{\mathrm{RRK}}=0$&$\textbf{0}$\tablefootmark{e}&$\textbf{0}$\tablefootmark{e}&2.3(-11)&$\textbf{7.0(-12)}$&$\textbf{4.5(-11)}$&7.2(-10)&1.0(-12)&3&0.413&4.292(05)&89.2\\
            $a_{\mathrm{RRK}}=0.05$&2.0(-11)&9.0(-12)&8.8(-11)&4.6(-11)&$\textbf{6.7(-10)}$&$\textbf{9.0(-12)}$&2.1(-12)&5&0.451&3.568(05)&25.4\\
            $a_{\mathrm{RRK}}=0.1$&6.9(-11)&2.1(-11)&$\textbf{4.6(-10)}$&6.5(-11)&$\textbf{1.1(-09)}$&$\textbf{3.0(-11)}$&4.0(-12)&4&0.448&3.590(05)&26.5\\
            $E_{\mathrm{b}}=0.3E_{\mathrm{d}}$&1.6(-10)&1.6(-11)&$\textbf{3.6(-10)}$&7.3(-11)&2.6(-10)&1.0(-12)&9.0(-12)&6&0.598&3.686(05)&32.3\\
            Shock at $T=10 \, \mathrm{K}$&&&&&&&&&&&\\
            $T_{\mathrm{peak}}=200 \, \mathrm{K}$, $t=10^2 \, \mathrm{yr}$&6.1(-11)&1.1(-11)&7.8(-11)&4.4(-11)&$\textbf{1.0(-09)}$&3.0(-12)&1.2(-11)&6&0.465&3.993(05)&56.8\\
            $T_{\mathrm{peak}}=1000 \, \mathrm{K}$, $t=10^2 \, \mathrm{yr}$&2.6(-10)&2.9(-11)&$\textbf{1.3(-10)}$&8.2(-11)&4.5(-11)&$\textbf{1.4(-11)}$&1.0(-12)&5&0.676&4.011(05)&58.5\\
            Shock at $T=20 \, \mathrm{K}$&&&&&&&&&&&\\
            $T_{\mathrm{peak}}=200 \, \mathrm{K}$, $t=10^2 \, \mathrm{yr}$&$\textbf{0}$\tablefootmark{e}&$\textbf{0}$\tablefootmark{e}&2.5(-11)&$\textbf{5.0(-12)}$&$\textbf{3.0(-09)}$&$\textbf{9.0(-12)}$&1.0(-12)&2&0.299&3.952(05)&53.0\\
            $T_{\mathrm{peak}}=1000 \, \mathrm{K}$, $t=10^2 \, \mathrm{yr}$&$\textbf{0}$\tablefootmark{e}&$\textbf{0}$\tablefootmark{e}&8.0(-11)&$\textbf{9.0(-12)}$&$\textbf{1.2(-09)}$&$\textbf{1.1(-11)}$&1.0(-12)&2&0.274&3.998(05)&57.3\\
            X-ray burst at $T=10 \, \mathrm{K}$&&&&&&&&&&&\\
            $\zeta_{\mathrm{CR}}=1.3\times10^{-14} \, \mathrm{s^{-1}}$, $t=10^2 \, \mathrm{yr}$&4.3(-11)&1.6(-11)&$\textbf{2.9(-10)}$&2.1(-11)&1.4(-10)&2.0(-12)&2.0(-12)&6&0.549&3.630(05)&28.8\\
            $\zeta_{\mathrm{CR}}=1.3\times10^{-14} \, \mathrm{s^{-1}}$, $t=10^3 \, \mathrm{yr}$&1.4(-10)&1.0(-11)&$\textbf{4.4(-10)}$&2.2(-10)&2.3(-10)&3.0(-12)&1.5(-11)&6&0.564&4.019(05)&59.3\\
            $\zeta_{\mathrm{CR}}=1.3\times10^{-13} \, \mathrm{s^{-1}}$, $t=10^2 \, \mathrm{yr}$&3.2(-11)&8.0(-12)&9.8(-11)&$\textbf{1.2(-11)}$&1.1(-10)&1.0(-12)&1.0(-12)&6&0.590&3.599(05)&27.0\\
            $\zeta_{\mathrm{CR}}=1.3\times10^{-13} \, \mathrm{s^{-1}}$, $t=10^3 \, \mathrm{yr}$&$\textbf{1.0(-12)}$&$\textbf{0}$\tablefootmark{e}&1.8(-11)&4.9(-11)&1.6(-11)&4.1(-13)&2.0(-12)&5&0.542&4.080(05)&65.4\\
            X-ray burst at $T=20 \, \mathrm{K}$&&&&&&&&&&&\\
            $\zeta_{\mathrm{CR}}=1.3\times10^{-14} \, \mathrm{s^{-1}}$, $t=10^2 \, \mathrm{yr}$&5.8(-11)&1.8(-11)&9.8(-11)&2.1(-11)&1.9(-10)&$\textbf{5.0(-12)}$&3.0(-12)&6&0.536&3.592(05)&26.6\\
            $\zeta_{\mathrm{CR}}=1.3\times10^{-14} \, \mathrm{s^{-1}}$, $t=10^3 \, \mathrm{yr}$&2.8(-11)&1.5(-11)&7.2(-11)&4.1(-11)&1.1(-10)&3.9(-12)&7.9(-13)&7&0.622&3.632(05)&28.9\\
            $\zeta_{\mathrm{CR}}=1.3\times10^{-13} \, \mathrm{s^{-1}}$, $t=10^2 \, \mathrm{yr}$&1.4(-11)&1.1(-11)&1.9(-11)&2.1(-11)&5.2(-11)&2.2(-12)&8.3(-13)&7&0.676&3.605(05)&27.4\\
            $\zeta_{\mathrm{CR}}=1.3\times10^{-13} \, \mathrm{s^{-1}}$, $t=10^3 \, \mathrm{yr}$&$\textbf{0}$\tablefootmark{e}&$\textbf{1.0(-12)}$&1.8(-11)&1.9(-10)&2.5(-11)&5.2(-13)&2.0(-12)&5&0.640&4.079(05)&65.3\\
          \hline
          \end{tabular} \\
          \tablefoot{
            \tablefoottext{a}{\citet{2020MNRAS.492..556L}. For five extended COMs the upper and lower limits of the abundance are provided; for two compact COMs only the upper limit of the abundance is provided.}
            \tablefoottext{b}{The median values are selected from the abundances of the observed points in the extended region of Sgr B2 to  calculate $\kappa$.}
            \tablefoottext{c}{$a(b)=a\times10^b .$ }
            \tablefoottext{d}{Bold  indicates overproduction compared with the upper limit or underproduction compared with the lower limit.}
            \tablefoottext{e}{These numbers are far below 10$^{-13}$ and for convenience are designated as 0.}
            \tablefoottext{f}{Number of fits represents how many COMs can fit observations at $t_{\mathrm{total}}$.}
            \tablefoottext{g}{$\kappa$ is the maximum average value of seven confidence levels in each model.}
            \tablefoottext{h}{$t_{\mathrm{total}}$ is the entire evolutionary time when $\kappa$ is the maximum value.}
            \tablefoottext{i}{$T$ is the temperature at $t_{\mathrm{total}}$.}
          }       
       \end{sidewaystable*}
       
   \par{}Table \ref{table3} shows the comparison of the simulations of several evolving models with the observations in the extended region of Sgr B2. For each species, if its abundance is higher than the upper limit or lower than the lower limit, the simulation of this species cannot fit the observation. We consider both the value of $\kappa$ and the number of fits to evaluate the best agreement of model with the data. 
   
   \par{}For fiducial model~2, two agreement values are presented, $\kappa=0.473$ when $T=36.8\, \mathrm{K}$ and $\kappa=0.565$ when $T=57.2\, \mathrm{K}$. In both these cases, observations of six COMs can be fitted well, except for t-HCOOH which remains overproduced. However, despite such a reasonable agreement, the best-fit dust temperatures are higher than the observed value of about 20~K. Other models also show two best-fit cases, when $T\sim30\, \mathrm{K}$ and $T\sim60\, \mathrm{K}$, respectively. Only the agreement of the maximum $\kappa$ is listed in Table \ref{table3}. Most of these models cannot fit the observations better than fiducial model~2.
   \par{}In the models with different ``usual'' key parameters ($n_{\mathrm{H}}$, $A_{\mathrm{V}}$, $t_{\mathrm{cold}}$, $t_{\mathrm{warmup}}$, $T_{\mathrm{max}}$, $G_0$, or $\zeta_{\mathrm{CR}}$), the number of fitted COM abundances can reach six in about half of the 17 models, while most values of $\kappa$ in these models are larger than that in fiducial model~2. Even so, the best-fit abundances in these models occur at $T\sim 50\, \mathrm{K}$, which is higher than the observed dust temperature. In the other half of these models, the number of fitted COM abundances is less than six, so they deviate from the observations farther than fiducial model~2. Thus, these 17 models cannot fit the observations better than fiducial model~2. In the models with $a_{\mathrm{RRK}}=0$, 0.05, 0.1, and $E_{\mathrm{b}}=0.3E_{\mathrm{d}}$, the overall agreement is as mediocre as it is in the17 models. Thus, adopting different key parameters cannot much improve the agreement with the observations at $T<30\, \mathrm{K}$. 
   \par{}In the models with a shock, if it happens early when $T=10 \, \mathrm{K}$, the agreement with observations also does not improve. If the shock occurs when $T=20 \, \mathrm{K}$, the results become worse and still cannot fit the observations because of the underproduction of CH$_2$OHCHO, CH$_3$OCHO, and C$_2$H$_5$OH, and the overproduction of CH$_3$NH$_2$ and CH$_3$OCH$_3$. Thus, an addition of a shock to the evolving model does not help to improve its feasibility. 
   \par{}Finally, the models with an early X-ray burst at $T=10 \, \mathrm{K}$ still cannot fit all seven COMs at the same time. Only the models with an X-ray burst at a later time moment, when $T=20 \, \mathrm{K}$ and the effective ionization rate becomes $\zeta_{\mathrm{CR}}=1.3\times10^{-14} \, \mathrm{s^{-1}}$, while the X-ray burst duration is $t=10^3 \, \mathrm{yr}$ or when $\zeta_{\mathrm{CR}}=1.3\times10^{-13} \, \mathrm{s^{-1}}$ and $t=10^2 \, \mathrm{yr}$ can fit all seven COMs at the same time moment when $T\sim28\, \mathrm{K}$. The corresponding agreement value $\kappa$ is 0.622 and 0.676, respectively. Thus, by considering an extra X-ray burst when $T=20 \, \mathrm{K}$ in the warm-up phase with a short timescale, a good fit to the COM observations in Sgr~B2 can be achieved. We conclude that in the framework of the present study, we achieve the best agreement with the observations with the evolving model with the following parameters: $n_{\mathrm{H}}=2\times10^3 \, \mathrm{cm^{-3}}$, $A_{\mathrm{V}}=10$, $t_{\mathrm{cold}}=3\times10^5 \, \mathrm{yr}$, $t_{\mathrm{warmup}}=2\times10^5 \, \mathrm{yr}$, $T_{\mathrm{max}}=200 \, \mathrm{K}$, $G_0=10^4$ with an X-ray burst at $T=20 \, \mathrm{K}$ ($t_{\mathrm{total}}=3.459\times10^5 \, \mathrm{yr}$), effective ionization rate $\zeta_{\mathrm{CR}}=1.3\times10^{-13} \, \mathrm{s^{-1}}$ and $t=10^2 \, \mathrm{yr}$.

  \section{Discussions} \label{section4}
  \subsection{Importance of reactive desorption} \label{section4.1}
  \par{}Since only the results in the warm-up phase in the evolving models can fit the observations, we need to figure out how these COMs are produced. In fiducial model~2, when $T\sim30-60\, \mathrm{K}$, the key mechanism for producing these COMs is the reactive desorption due to the exothermic reactions on the grain surfaces. The most important reactions are listed below: 
        \begin{align}
        &\mathrm{JCH_2OH + JHCO \rightarrow CH_2OHCHO, }  \label{equation5} \\
        &\mathrm{JCH_3O + JHCO \rightarrow CH_3OCHO, }  \label{equation6} \\
        &\mathrm{JOH + JHCO \rightarrow HCOOH, }  \label{equation7} \\
        &\mathrm{JCH_2OH + JCH_3 \rightarrow C_2H_5OH, }  \label{equation8} \\
        &\mathrm{JC_2H_5O + JH \rightarrow C_2H_5OH, }  \label{equation9} \\
        &\mathrm{JCH_2NH_2 + JH \rightarrow CH_3NH_2, }  \label{equation10} \\
        &\mathrm{JCH_3NH + JH \rightarrow CH_3NH_2, }  \label{equation11} \\
        &\mathrm{JNH_2 + JCH_3 \rightarrow CH_3NH_2, }  \label{equation12} \\
        &\mathrm{JCH_3O + JCH_3 \rightarrow CH_3OCH_3, }  \label{equation13} \\
        &\mathrm{JC_2H_5 + JCN \rightarrow C_2H_5CN, }  \label{equation14} \\
        &\mathrm{JCH_2CN + JCH_3 \rightarrow C_2H_5CN, }  \label{equation15} \\
        &\mathrm{JC_2H_4CN + JH \rightarrow C_2H_5CN, }  \label{equation16} \\
        &\mathrm{JCH_3CHCN + JH \rightarrow C_2H_5CN. }  \label{equation17} 
        \end{align}
        Here the letter J represents species on the grain. These COMs are mainly destroyed by the ion-molecule reactions in the gas which are listed below: 
        \begin{align}
        &\mathrm{CH_2OHCHO + C^+ \rightarrow CH_2OHCHO^+ + C, }  \label{equation18} \\
        &\mathrm{CH_3OCHO + C^+ \rightarrow CH_3OCHO^+ + C, }  \label{equation19} \\
        &\mathrm{HCOOH + C^+ \rightarrow HCOOH^+ + C, }  \label{equation20} \\
        &\mathrm{C_2H_5OH + C^+ \rightarrow C_2H_5OH^+ + C, }  \label{equation21} \\
        &\mathrm{C_2H_5OH + C^+ \rightarrow CH_2OH^+ + C_2H_3, }  \label{equation22} \\
        &\mathrm{C_2H_5OH + C^+ \rightarrow CH_3CH_2O^+ + CH, }  \label{equation23} \\
        &\mathrm{CH_3NH_2 + C^+ \rightarrow CH_3NH_2^+ + C, }  \label{equation24} \\
        &\mathrm{CH_3NH_2 + C^+ \rightarrow CH_3NH^+ + CH, }  \label{equation25} \\
        &\mathrm{CH_3OCH_3 + C^+ \rightarrow CH_3OCH_3^+ + C, }  \label{equation26} \\
        &\mathrm{C_2H_5CN + C^+ \rightarrow C_2H_5CN^+ + C, }  \label{equation27} \\
        &\mathrm{C_2H_5CN + He^+ \rightarrow CH_3^+ + CH_2 + CN + He. }  \label{equation28} 
        \end{align}
        Here C$^+$ is the key species that destroys them. The products of these reactions react with $e^-$ and produce multiple smaller molecules, which cannot be re-assembled into COMs in the gas phase. The relatively high abundances of CH$_2$OHCHO, CH$_3$OCHO, t-HCOOH, C$_2$H$_5$OH, and CH$_3$NH$_2$ and low abundances of CH$_3$OCH$_3$ and C$_2$H$_5$CN are mainly determined by the the reactive desorption. 
        
        \par{}The importance of the reactive desorption for the gas-phase abundances of the seven COMs can be explained as follows. Reactions 5-17 belong to the Langmuir-Hinshelwood formation mechanism, and the reaction rates are mainly determined by the diffusion rates of radicals, especially the ones with lower $E_{\mathrm{b}}$ (or $E_{\mathrm{d}}$). The evaporation temperatures of JHCO, JCH$_3$, JH, and JCN are about 26 K, 20 K, 8 K, and 26 K, respectively, according to their desorption energies $E_{\mathrm{d}}$. Reactions 5-7 should be highly efficient when $T\sim26\, \mathrm{K}$ because of the high diffusion rate of JHCO. They produce CH$_2$OHCHO, CH$_3$OCHO, and t-HCOOH rapidly. The abundance of JOH is higher than that of JCH$_2$OH and JCH$_3$O, because it is easier for JH to react with JO and produce JOH than to react with JCO, JHCO, and JCH$_2$O and produce more complex ices JCH$_2$OH and JCH$_3$O. This leads to the overall higher abundance of t-HCOOH compared to the other six larger COMs. JCH$_3$OH can be photodissociated by the UV photons to produce JCH$_2$OH and JCH$_3$O, but the efficiency for producing JCH$_2$OH is five times higher than the efficiency for producing JCH$_3$O. Thus, the abundance of JCH$_2$OH is larger than JCH$_3$O and results in more abundant CH$_2$OHCHO and less abundant CH$_3$OCHO molecules. Similarly, the increase in C$_2$H$_5$OH and CH$_3$OCH$_3$ abundances when $T\sim20\, \mathrm{K}$ is associated with the high diffusion rate of JCH$_3$ in reactions 8 and 13 and with the higher abundance of JCH$_2$OH leads to more C$_2$H$_5$OH molecules, while the lower abundance of JCH$_3$O results in fewer CH$_3$OCH$_3$ molecules. In addition, reaction 9 can also increase the abundance of C$_2$H$_5$OH slightly. According to reactions 6 and 13, the lower abundance of JCH$_3$ radicals compared to JHCO radicals leads to the lower abundance of CH$_3$OCH$_3$ compared to CH$_3$OCHO, which also makes the abundance of CH$_3$OCH$_3$ the lowest of all seven COMs. For CH$_3$NH$_2$, reactions 10 and 11 can be efficient when $T\sim8\, \mathrm{K}$, so CH$_3$NH$_2$ can be produced in the cold phase and hence also in the static models. However, the low abundances of JCH$_2$NH$_2$ and JCH$_3$NH molecules prevent CH$_3$NH$_2$ from being produced efficiently. When $T\sim20\, \mathrm{K}$, the high diffusion rate of JCH$_3$ in reaction 12 leads to the increase in abundance of CH$_3$NH$_2$. For C$_2$H$_5$CN reaction 14 is significant if $T\sim26\, \mathrm{K}$, when the high diffusion rate of JCN leads to the increased production of C$_2$H$_5$CN. In addition to the ion-molecule reactions with C$^+$, reaction 28 can decrease the abundance of C$_2$H$_5$CN slightly to maintain its relatively low abundance. Therefore, the relatively high abundances of CH$_2$OHCHO, CH$_3$OCHO, t-HCOOH, C$_2$H$_5$OH, and CH$_3$NH$_2$ and low abundances of CH$_3$OCH$_3$ and C$_2$H$_5$CN can be obtained when $T\sim30-40\, \mathrm{K}$. 
        \par{}Since CH$_2$OHCHO, CH$_3$OCHO, and t-HCOOH are produced efficiently after the temperature exceeds about 26~K, it is difficult to fit the observations of COMs when the measured dust temperature is lower ($\sim 20$~K). At higher temperatures ($T>60\, \mathrm{K}$), t-HCOOH, C$_2$H$_5$OH, CH$_3$NH$_2$, CH$_3$OCH$_3$, and C$_2$H$_5$CN abundances reach their peak abundances due to thermal evaporation. However, CH$_2$OHCHO and CH$_3$OCHO abundances decrease because the radical JOH begins to react with JCH$_2$OHCHO and JCH$_3$OCHO when $T\sim50\, \mathrm{K}$, which leads to their destruction and prevents their evaporation from increasing the abundances in the gas-phase. In addition, the ion-molecule reactions in the gas-phase destroy CH$_2$OHCHO and CH$_3$OCHO molecules to decrease their abundances. These reactions also become important when the temperature is high, so all seven abundances decrease at the end of the warm-up phase. Thus, the COM abundances representative of the observations are obtained when $T\sim30-60\, \mathrm{K}$ in the models without an extra X-ray burst.

        \par{}It is clear from the above analysis of the importance of reactive desorption that the simulations of the static models cannot fit the observations. If $T_{\mathrm{dust}}=20\, \mathrm{K}$,  then JCO and JCH$_3$ evaporate, while more complex radicals (JC$_2$OH, JCH$_3$O, JHCO, and JC$_2$H$_5$O) are not produced efficiently, such that the abundances of CH$_2$OHCHO, CH$_3$OCHO, t-HCOOH, C$_2$H$_5$OH, and CH$_3$OCH$_3$ remain very low. Only CH$_3$NH$_2$ and C$_2$H$_5$CN can be produced through reactions 10, 11, and 14 and become abundant in the gas phase due to reactive desorption. Considering other mechanisms cannot solve the problem fundamentally. 
        
        \par{}In the evolving models, the parameters that can enhance the reaction rates of reactions 5-17 may result in better fits compared to the observations. This explains why the impact of different values of $n_{\mathrm{H}}$, $t_{\mathrm{cold}}$, $t_{\mathrm{warmup}}$, and $T_{\mathrm{max}}$ is negligible. The value of $A_{\mathrm{V}}$ must be larger than 10~mag, otherwise these COMs and their large precursor radicals are too strongly photodissociated by the UV photons. At the dark conditions with $A_{\mathrm{V}}\ge10$, the unattenuated FUV intensity $G_0$ is not that important to the simulations if it is lower than $\sim 10^4$. Similarly, the cosmic ray ionization rate $\zeta_{\mathrm{CR}}$ should be less than $\sim 10^{-15} \, \mathrm{s}^{-1}$ on long timescales to avoid from destroying these COMs by cosmic ray ionization and dissociation, which is consistent with the derived values from chemical simulations for the hot cores in Sgr~B2(N) \citep{2019A&A...628A..27B, 2020A&A...636A..29W}. The larger value of $a_{\mathrm{RRK}}$ (0.05 or 0.1) allow more COMs to reach the gas phase, but CH$_3$NH$_2$ and CH$_3$OCH$_3$ abundances are overproduced. Thus, $a_{\mathrm{RRK}}=0.01$ is a reasonable value. The lower $E_{\mathrm{b}}$ can  enhance the diffusion rates, leading to more efficient surface chemistry, higher COM ice abundances and a better agreement with the observations. The diffusion and binding energies for complex ices are poorly known and require more laboratory studies to help to improve the results of astrochemical models, including this study. 
        \par{}During a shock passage, gas and dust temperatures can temporarily reach high values. If it happens when $T=10\, \mathrm{K}$, all ices on the grain evaporate, shortening the surface chemistry efficiency. Most of the complex molecules are destroyed by ion-molecule reactions before freezing out back to the grain surfaces. Most radicals are adsorbed on the grain surface intact after the shock passage, but the COMs produced from them later cannot easily reach the gas phase. However, if a shock happens at a later time when $T=20\, \mathrm{K}$, t-HCOOH, C$_2$H$_5$OH, CH$_3$NH$_2$, CH$_3$OCH$_3$, and C$_2$H$_5$CN abundances increase rapidly during its passage. The last four can be produced efficiently through reactions 8, 12, 13, and 15 on the grains and then become evaporated during the shock, while t-HCOOH can be partly produced through the gas-phase reactions. CH$_2$OHCHO and CH$_3$OCHO take longer to produce in sufficient amounts on the dust surfaces, and the lack of JCH$_2$OH, JCH$_3$O, and JHCO after the shock passage blocks production of these two COMs through reactions 5 and 6. This explains why models with a shock cannot explain the observations. 
        \par{}An X-ray burst can enhance ionization and dissociation reactions, like those produced by cosmic rays, and can increase non-thermal desorption. If it happens early, when $T=10\, \mathrm{K}$, its impact can be ignored since surface chemistry has no time to produce abundant COMs ices. If it happens later, when $T=20\, \mathrm{K}$, the photodesorption induced by the X-rays increases the gas-phase abundances of all COMs immediately, but later leads to their destruction in the gas via ion-molecule reactions. This mechanism leads to peak COM gas-phase abundances that are achieved shortly after the onset of the burst, and they decrease afterwards. After the X-ray burst ends, the COM abundances decrease slightly when $T\sim30\, \mathrm{K}$. The reason is that the radicals on the grain surfaces producing these COMs become dissociated during the X-ray burst, which thus suppresses the production of COMs at later times. This is the scenario that shows the best agreement between our models and the Sgr~B2 observations.

  \subsection{Origin of the chemical differentiation of COMs} \label{section4.2}
  \par{}The X-ray flare released by the supermassive black hole Sgr A* at the center of the Milky Way has been observed \citep{2001Natur.413...45B, 2008ApJ...682..373M, 2013ApJ...775L..34R, 2017MNRAS.465...45C}, presumably created by the accretion of gas onto the black hole. It probably occurred about 100~years ago \citep{2017MNRAS.465...45C}. Such a flare could be multiple, shorter-duration flares superposed on a long-term high state, since two short-term flares of 5-10 years have been found \citep{2013PASJ...65...33R}. The X-ray echo from Sgr~B2 has also been observed \citep{2010ApJ...719..143T, 2011ApJ...739L..52N}, which indicates that the region around Sgr~B2 can be continuously affected by the X-ray flares from Sgr A*. The abundances of two organic molecules CH$_3$OH and H$_2$CO can be enhanced by the X-ray irradiation from Sgr A*, which has been simulated by \citet{2020ApJ...899...92L}. Combining our simulations with the observations, a possible explanation for the chemical differentiation in the extended region around Sgr~B2 is proposed.

        \begin{figure}
        \begin{center}
           \includegraphics[width=8cm]{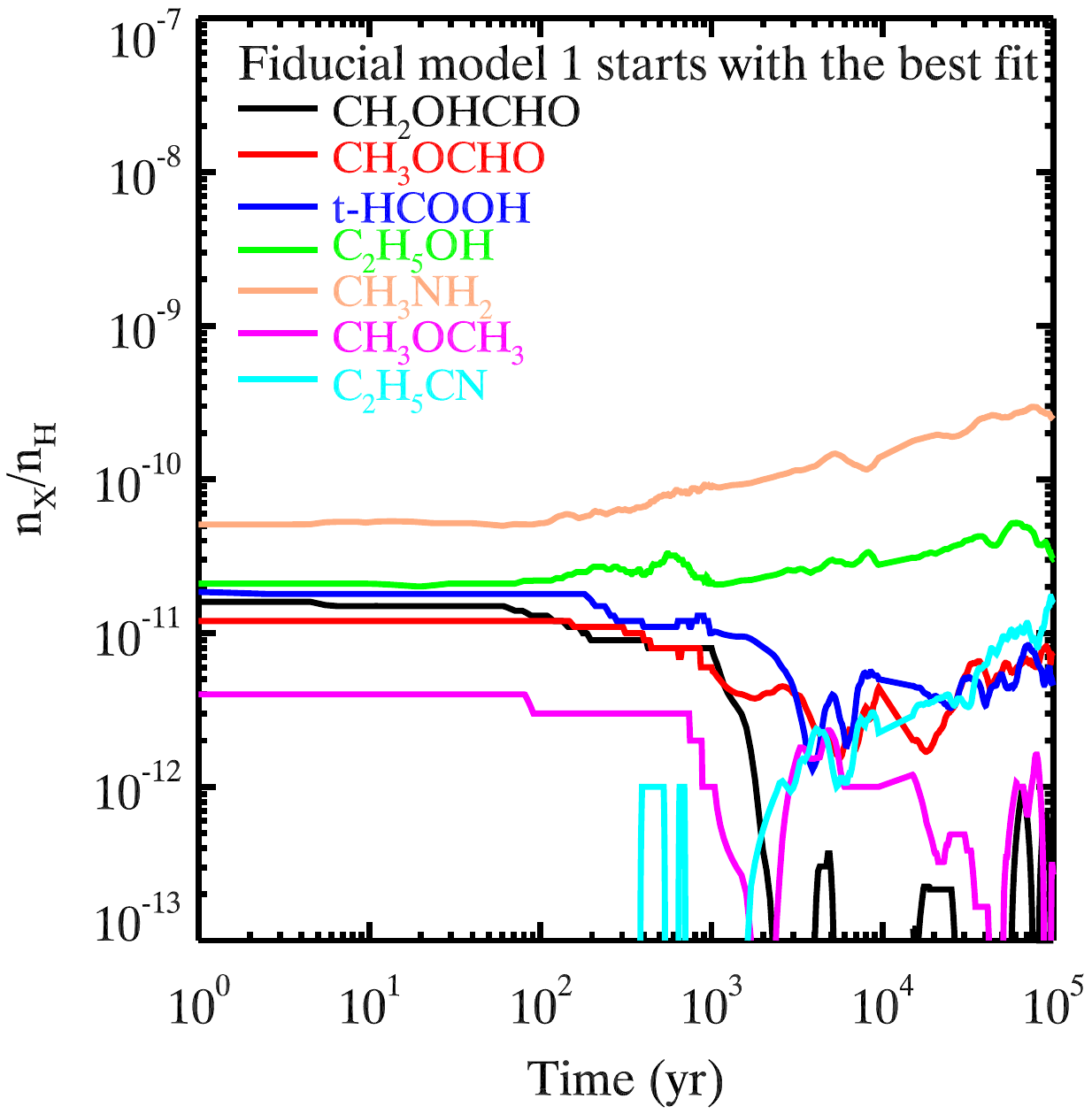}
           \end{center}
            \caption{Time-dependent evolution of the relative gas-phase abundances of CH$_2$OHCHO, CH$_3$OCHO, t-HCOOH, C$_2$H$_5$OH, CH$_3$NH$_2$, CH$_3$OCH$_3$, and C$_2$H$_5$CN in fiducial model~1, when adopting the abundances at $t_{\mathrm{total}}=3.605\times10^5\, \mathrm{yr}$ from fiducial model~2 with an X-ray burst at $T=20 \, \mathrm{K}$ with $\zeta_{\mathrm{CR}}=1.3\times10^{-13} \, \mathrm{s^{-1}}$ and $t=10^2 \, \mathrm{yr}$ as the initial abundances. }
            \label{figure5}
        \end{figure}

  \par{}The extended region around Sgr~B2 underwent a cold phase with $T\le10\, \mathrm{K}$ for $10^5-10^6\, \mathrm{yr}$ and then a warm-up phase began due to the star formation. When the temperature reached roughly $T\sim20\, \mathrm{K}$, a short-term X-ray flare lasting no more than 100 years from Sgr A* occurred and affected the region. It enhanced the high-energy ionization and dissociation rates, as well as high-energy-driven desorption and induced photodesorption, while simultaneously heating the gas up to  temperature of 65~K. The X-ray and UV photons produced near the Galactic center would be the major heating agent. After the flare ended, the dust temperature continued to rise to about 27~K because of the star formation. At that moment the abundances of the seven COMs were obtained. The timescale for this period is variable and depends on the warm-up mechanism. After that, the gas and dust temperature could continue to rise. Under such circumstances, the observed COM abundances can be sustained at $T\sim27-50\, \mathrm{K}$, which lasts for about $3\times10^4$ years in our fiducial model~2. An alternative is that the gas and dust temperature remain 65~K and 20~K, respectively. However, the best fit cannot be sustained for a long time. We reran static fiducial model~1 by adopting the results when the best fit occurs according to Table \ref{table3} as the initial abundances, so the abundances at $t_{\mathrm{total}}=3.605\times10^5\, \mathrm{yr}$ in fiducial model~2 with an X-ray burst at $T=20 \, \mathrm{K}$, $\zeta_{\mathrm{CR}}=1.3\times10^{-13} \, \mathrm{s^{-1}}$ and $t=10^2 \, \mathrm{yr}$ are selected. Figure~\ref{figure5} depicts the results. The best fit for all seven COMs can remain for only several hundred years, which seems to be consistent with the observed flare that could have happened about 100 years ago \citep{2017MNRAS.465...45C}. If the observed abundances need to be sustained longer, multiple short-term flares may have to be considered. 
  
  \par{}In addition to these seven COMs, we also check the chemical differentiation of other organic molecules when $T\sim30\, \mathrm{K}$. Some organic molecules are similar to five extended COMs and can reach relatively high abundances including CH$_3$CCH, HC$_3$N, HC$_5$N, CH$_3$OH, CH$_3$CN, H$_2$CO, CH$_3$CHO, NH$_2$CHO, and c-H$_2$C$_3$O. Some of them have been observed in the extended region around Sgr~B2 \citep{2006ApJ...642..933H, 2008MNRAS.386..117J, 2011MNRAS.411.2293J}. In contrast, some COMs show very low abundances including C$_2$H$_5$CHO, CH$_2$OHCH$_2$OH, HCOOCH$_2$CH$_3$, CH$_3$COOH, C$_2$H$_5$COOH, C$_2$H$_3$CN, C$_3$H$_7$CN, CH$_3$CONH$_2$, NH$_2$CH$_2$CN, and NH$_2$CH$_2$COOH, and they may not be observable in the extended region around Sgr~B2 just like CH$_3$OCH$_3$ and C$_2$H$_5$CN. However, C$_2$H$_5$CHO and CH$_3$CONH$_2$ have actually been detected in Sgr~B2 \citep{2004ApJ...610L..21H, 2006ApJ...643L..25H}, which is at odds with our model. Other detected species such as C$_2$H$_3$CHO, CNCHO, CH$_3$CHNH, and HNCHCN \citep{2004ApJ...610L..21H, 2008ApJ...675L..85R, 2013ApJ...765L...9L, 2013ApJ...765L..10Z} are not included in our reaction network yet. It indicates that our chemical models still can be improved, including additional reactions related to these species, and more studies about the chemical mechanisms and reactions for producing COMs at low temperature are needed. More observations searching for COMs not only in this region but also in the cold regions around other star-forming regions can help to verify the validity of our simulations.

  \section{Conclusions} \label{section5}
  \par{}Based on the known physical parameters in the extended region around Sgr~B2, we calculate a series of models to explore under what physical conditions the chemical simulations can fit the observations with the relatively high abundances of CH$_2$OHCHO, CH$_3$OCHO, t-HCOOH, C$_2$H$_5$OH, and CH$_3$NH$_2$, but the low abundances of CH$_3$OCH$_3$ and C$_2$H$_5$CN. We also try to explain the deficiency of CH$_3$OCH$_3$ and C$_2$H$_5$CN. The macroscopic Monte Carlo method is used in the simulations. Photodesorption is included in all models. We study the influences of the chain reaction mechanism, shock action, X-ray burst, enhanced reactive desorption, and low diffusion barriers in some models. We adopt two types of simple physical models, a static model and an evolving model that includes a cold phase and a warm-up phase. The fiducial models adopt the observed physical parameters except for an ill-defined local cosmic ray ionization rate $\zeta_{\mathrm{CR}}$, and we simulate a series of models by changing the value of one parameter each time. All static models are not able to fit the observations, and in most models only the abundances of CH$_3$NH$_2$ and C$_2$H$_5$CN can become relatively high. Most evolving models can fit at most six out of the seven COMs for $T \sim 30-60 \, \mathrm{K}$ in the warm-up phase, but the best-fit temperature is still higher than the observed dust temperature of 20~K. The best agreement between the simulations and all seven observed COMs at a lower temperature $T\sim 27 \, \mathrm{K}$ is achieved by considering a short-duration, $\approx 10^2$~yr X-ray burst with $\zeta_{\mathrm{CR}}=1.3 \times 10^{-13} \, \mathrm{s}^{-1}$ at the early stage of the warm-up phase, when it still has a temperature of $20 \, \mathrm{K}$. The reactive desorption is the key mechanism for producing these COMs and inducing the low abundances of CH$_3$OCH$_3$ and C$_2$H$_5$CN, because the radicals on the grain for producing these two COMs are less abundant than the radicals for producing the other five COMs. A possible explanation for the chemical differentiation is that the extended region around Sgr~B2 underwent a cold phase with $T\le10\, \mathrm{K}$ and then a warm-up phase, when $T\sim20\, \mathrm{K}$ an X-ray flare from a Galactic black hole Sgr~A* with a short duration of no more than 100 years was acquired, affecting strongly the Sgr~B2 chemistry. The observed COM abundances in Sgr~B2 can remain only several hundred years after such a flare, which could imply that such short-term X-rays flares occur relatively often, likely associated with the accretion activity of the Sgr~A* source. \\

  \begin{acknowledgements}
We thank Robin T. Garrod for providing the chemical reaction network he uses for the simulations of COMs. Y.W. thanks Qiang Chang for the suggestions of the Monte Carlo method and Chong Li for helping calculate some column densities. Y.W. and H.W. thank the supports by the National Key R\&D Program of China (No. 2017YFA0402701) and by National Natural Science Foundation of China (NSFC) grant No. 11973091. Y.W. acknowledges the support from China Scholarship Council (CSC) grant No. 201906340047. F.D. is supported by the National Natural Science Foundation of China (NSFC) grant No. 11873094 and 12041305. D.S. acknowledges support by the Deutsche Forschungsgemeinschaft through SPP 1833: ``Building a Habitable Earth'' (SE 1962/6-1). This research made use of NASA's Astrophysics Data System. 
  \end{acknowledgements}
  
  \bibliographystyle{aa} 
  \bibliography{references}

\begin{thebibliography}{74}
\expandafter\ifx\csname natexlab\endcsname\relax\def\natexlab#1{#1}\fi

\bibitem[{{Bacmann} {et~al.}(2012){Bacmann}, {Taquet}, {Faure}, {Kahane}, \&
  {Ceccarelli}}]{2012A&A...541L..12B}
{Bacmann}, A., {Taquet}, V., {Faure}, A., {Kahane}, C., \& {Ceccarelli}, C.
  2012,
  \href{https://www.aanda.org/articles/aa/pdf/2012/05/aa19207-12.pdf}{\aap},
  541, L12

\bibitem[{{Baganoff} {et~al.}(2001){Baganoff}, {Bautz}, {Brandt}, {Chartas},
  {Feigelson}, {Garmire}, {Maeda}, {Morris}, {Ricker}, {Townsley}, \&
  {Walter}}]{2001Natur.413...45B}
{Baganoff}, F.~K., {Bautz}, M.~W., {Brandt}, W.~N., {et~al.} 2001,
  \href{https://www.nature.com/articles/35092510}{\nat}, 413, 45

\bibitem[{{Balucani} {et~al.}(2015){Balucani}, {Ceccarelli}, \&
  {Taquet}}]{2015MNRAS.449L..16B}
{Balucani}, N., {Ceccarelli}, C., \& {Taquet}, V. 2015,
  \href{https://watermark.silverchair.com/slv009.pdf?token=AQECAHi208BE49Ooan9kkhW_Ercy7Dm3ZL_9Cf3qfKAc485ysgAAArkwggK1BgkqhkiG9w0BBwagggKmMIICogIBADCCApsGCSqGSIb3DQEHATAeBglghkgBZQMEAS4wEQQMGtIZilfWZ7vEixnmAgEQgIICbIZl1IEWQAy_TolRRH9uQARP7EKIevJ9hL4SVj8HO6_QIzpjSo7RaJpJaV48ZAX2CXkwBbP-8A-gPqOv_sa9xo1kOiWcQpolsCLOYmvAHuRYH8nBDQIkptGgmwOf0gC-3Fyd1jov9k_534lmw0fWF_AaiQ395tlSiCj-ZE8YmXYn4cSHhKdJRzbOum3_tA4H1qL65r7YYvbtvL8RKueczZNhXSOUs4XZW65LWKkJEoMd3T7lrYE-4XAUwK7b_3AQt80nHuFRSOlOpiGip7yhoKWzHv-vRPObol06DO7UA2am6mNJx4Nlh70xa3mLt4115SC-TcLkpnYcXdvGEuMq--Dcta_7QPz1gGBsSd6wkZTGlnQe2Ooqd-kF5un1k3-EUT6NuTvk5sRhBIz7iuTxA7JefVA23_s7shQXjIShtm8ocupZYhQs49ESr_9rDUTucYk0skSCO9xYokOB60gLb9Aa9DNCnlG447t_8HrHBnhA8i3bAbNdiKxMOFS5XPBjFOaOTevehUL_GDBxezhSlfCFoYSiwFNyfF964Ov7ldd_P-V_Cibzt7DVi677ZW9jjIkv00NU9dDrdjJW5kS4mSq-GmLVj6SYx4qtXZqzCl-K-LdVYgZzrzJmGyqW7xapXqWAb-Zb_2BpOIxI8UIWeXQgaJaiNywLEFUXd-jSwcS1kMIqAwcm7BZ-AMjeERxnML8PUTsM3JP6TlNaA4RtZIiWBBxIkt1Q9xa8fj67Q6CzGMVd-PDBmmz-kAMJL9dpOWTNo9c5SNEeGJWFkgBzZPl5rNoCoiDV9Zr8C_oKBgjiGRkQ8bytaZi2jlkk}{\mnras},
  449, L16

\bibitem[{{Belloche} {et~al.}(2014){Belloche}, {Garrod}, {M{\"u}ller}, \&
  {Menten}}]{2014Sci...345.1584B}
{Belloche}, A., {Garrod}, R.~T., {M{\"u}ller}, H. S.~P., \& {Menten}, K.~M.
  2014,
  \href{https://science.sciencemag.org/content/345/6204/1584.full}{Science},
  345, 1584

\bibitem[{{Belloche} {et~al.}(2017){Belloche}, {Meshcheryakov}, {Garrod},
  {Ilyushin}, {Alekseev}, {Motiyenko}, {Margul{\`e}s}, {M{\"u}ller}, \&
  {Menten}}]{2017A&A...601A..49B}
{Belloche}, A., {Meshcheryakov}, A.~A., {Garrod}, R.~T., {et~al.} 2017,
  \href{https://www.aanda.org/articles/aa/pdf/2017/05/aa29724-16.pdf}{\aap},
  601, A49

\bibitem[{{Belloche} {et~al.}(2013){Belloche}, {M{\"u}ller}, {Menten},
  {Schilke}, \& {Comito}}]{2013A&A...559A..47B}
{Belloche}, A., {M{\"u}ller}, H.~S.~P., {Menten}, K.~M., {Schilke}, P., \&
  {Comito}, C. 2013,
  \href{https://www.aanda.org/articles/aa/pdf/2013/11/aa21096-13.pdf}{\aap},
  559, A47

\bibitem[{{Bergner} {et~al.}(2017){Bergner}, {{\"O}berg}, {Garrod}, \&
  {Graninger}}]{2017ApJ...841..120B}
{Bergner}, J.~B., {{\"O}berg}, K.~I., {Garrod}, R.~T., \& {Graninger}, D.~M.
  2017,
  \href{https://iopscience.iop.org/article/10.3847/1538-4357/aa72f6/pdf}{\apj},
  841, 120

\bibitem[{{Bonfand} {et~al.}(2019){Bonfand}, {Belloche}, {Garrod}, {Menten},
  {Willis}, {St{\'e}phan}, \& {M{\"u}ller}}]{2019A&A...628A..27B}
{Bonfand}, M., {Belloche}, A., {Garrod}, R.~T., {et~al.} 2019,
  \href{https://www.aanda.org/articles/aa/pdf/2019/08/aa35523-19.pdf}{\aap},
  628, A27

\bibitem[{{Burkhardt} {et~al.}(2019){Burkhardt}, {Shingledecker}, {Le Gal},
  {McGuire}, {Remijan}, \& {Herbst}}]{2019ApJ...881...32B}
{Burkhardt}, A.~M., {Shingledecker}, C.~N., {Le Gal}, R., {et~al.} 2019,
  \href{https://iopscience.iop.org/article/10.3847/1538-4357/ab2be8/pdf}{\apj},
  881, 32

\bibitem[{{Cernicharo} {et~al.}(2012){Cernicharo}, {Marcelino}, {Roueff},
  {Gerin}, {Jim{\'e}nez-Escobar}, \& {Mu{\~n}oz Caro}}]{2012ApJ...759L..43C}
{Cernicharo}, J., {Marcelino}, N., {Roueff}, E., {et~al.} 2012,
  \href{https://iopscience.iop.org/article/10.1088/2041-8205/759/2/L43/pdf}{\apjl},
  759, L43

\bibitem[{{Chang} \& {Herbst}(2012)}]{2012ApJ...759..147C}
{Chang}, Q. \& {Herbst}, E. 2012,
  \href{http://iopscience.iop.org/article/10.1088/0004-637X/759/2/147/pdf}{\apj},
  759, 147

\bibitem[{{Chang} \& {Herbst}(2014)}]{2014ApJ...787..135C}
{Chang}, Q. \& {Herbst}, E. 2014,
  \href{http://iopscience.iop.org/article/10.1088/0004-637X/787/2/135/pdf}{\apj},
  787, 135

\bibitem[{{Chang} \& {Herbst}(2016)}]{2016ApJ...819..145C}
{Chang}, Q. \& {Herbst}, E. 2016,
  \href{https://iopscience.iop.org/article/10.3847/0004-637X/819/2/145/pdf}{\apj},
  819, 145

\bibitem[{{Churazov} {et~al.}(2017){Churazov}, {Khabibullin}, {Sunyaev}, \&
  {Ponti}}]{2017MNRAS.465...45C}
{Churazov}, E., {Khabibullin}, I., {Sunyaev}, R., \& {Ponti}, G. 2017,
  \href{https://watermark.silverchair.com/stw2750.pdf?token=AQECAHi208BE49Ooan9kkhW_Ercy7Dm3ZL_9Cf3qfKAc485ysgAAArgwggK0BgkqhkiG9w0BBwagggKlMIICoQIBADCCApoGCSqGSIb3DQEHATAeBglghkgBZQMEAS4wEQQMX7FTaT0b00_WBWK0AgEQgIICa3t7o7ZNz4RYErvkvCrPgRWTTY1KQqmez8WGl-g3FqQ4RHCwzh2tjKt16g1jycNyC74O-ib5qn0sTK6zqv9Hz9ryAzYhmpLhcuOUaPeNXaZvHtOvzlMsE8qXddldnVNcKOaVeOSacQfBwfR7vzQFQHJqYAWqaslVSFwf8J8qinwnt9T4e4P5Rs2OLtx6nPf2VGTPejHpWBaY5JDdW3JC4oIHVjLtPWDazAY5HwJ_nyYpX7rczk1hbrqyWyOiV-g5iBQGEGVVJklYHZcC7zp4IbsY3-Cw9yKAQ5sTL4qWhX2p6w9TTDV0ngSifllgT8lnUU0f_4PqEWpRe_0xcZnc84PztEzhLyUDXaD3afM03BlNA8lWoRDbsO3Xm6d9XDLNGcQLL3cB6AWDUSQxseb0K0pJetlnGlBZUjN00Ejs76UV-qj9WDoBN38zJ1B9JnheRgj5EjWbHu8Zeah5LqK8fmqKUX2Rdkg-D2ndxjTS2pqutpFYpZYnnzHnJy0KZ9pfTWsiPRPuJabQt5muRBKmanHfI-JTAm91EuD2cNw5xGvDxzbX4CXvLXe3t4lAhK49gmZvunplgXOCG-6ZKK_8xADjf_G2ydQ9ifTZ58RDlmyPEjYVBJjV4FwO4a-5ZGtI_EICkVvDjp914FdkgRmNuS-3jQmRuQKwT6S0j4rQcLd2JmAlSSKKdpqn0s9n0NWkIJvnlfHmwwrm2hqeDnvjQeCBWnKJyCkvRDetNYj3BDIu-2MW-3jzRitAv5wOrgSyCNtZNbGnjViPaFo4ZCEWUCar3yPjsCb71cJPtFXxg3mglA-5PDw-mBkEAaw}{\mnras},
  465, 45

\bibitem[{{Cuppen} {et~al.}(2017){Cuppen}, {Walsh}, {Lamberts}, {Semenov},
  {Garrod}, {Penteado}, \& {Ioppolo}}]{2017SSRv..212....1C}
{Cuppen}, H.~M., {Walsh}, C., {Lamberts}, T., {et~al.} 2017,
  \href{https://link.springer.com/article/10.1007/s11214-016-0319-3}{\ssr},
  212, 1

\bibitem[{{Etxaluze} {et~al.}(2013){Etxaluze}, {Goicoechea}, {Cernicharo},
  {Polehampton}, {Noriega-Crespo}, {Molinari}, {Swinyard}, {Wu}, \&
  {Bally}}]{2013A&A...556A.137E}
{Etxaluze}, M., {Goicoechea}, J.~R., {Cernicharo}, J., {et~al.} 2013,
  \href{https://www.aanda.org/articles/aa/pdf/2013/08/aa21258-13.pdf}{\aap},
  556, A137

\bibitem[{{Garrod}(2013)}]{2013ApJ...765...60G}
{Garrod}, R.~T. 2013,
  \href{http://iopscience.iop.org/article/10.1088/0004-637X/765/1/60/pdf}{\apj},
  765, 60

\bibitem[{{Garrod}(2019)}]{2019ApJ...884...69G}
{Garrod}, R.~T. 2019,
  \href{https://iopscience.iop.org/article/10.3847/1538-4357/ab418e/pdf}{\apj},
  884, 69

\bibitem[{{Garrod} {et~al.}(2017){Garrod}, {Belloche}, {M{\"u}ller}, \&
  {Menten}}]{2017A&A...601A..48G}
{Garrod}, R.~T., {Belloche}, A., {M{\"u}ller}, H.~S.~P., \& {Menten}, K.~M.
  2017,
  \href{https://www.aanda.org/articles/aa/pdf/2017/05/aa30254-16.pdf}{\aap},
  601, A48

\bibitem[{{Garrod} \& {Herbst}(2006)}]{2006A&A...457..927G}
{Garrod}, R.~T. \& {Herbst}, E. 2006,
  \href{https://www.aanda.org/articles/aa/pdf/2006/39/aa5560-06.pdf}{\aap},
  457, 927

\bibitem[{{Garrod} {et~al.}(2007){Garrod}, {Wakelam}, \&
  {Herbst}}]{2007A&A...467.1103G}
{Garrod}, R.~T., {Wakelam}, V., \& {Herbst}, E. 2007,
  \href{https://www.aanda.org/articles/aa/pdf/2007/21/aa6704-06.pdf}{\aap},
  467, 1103

\bibitem[{{Garrod} {et~al.}(2008){Garrod}, {Widicus Weaver}, \&
  {Herbst}}]{2008ApJ...682..283G}
{Garrod}, R.~T., {Widicus Weaver}, S.~L., \& {Herbst}, E. 2008,
  \href{https://iopscience.iop.org/article/10.1086/588035/pdf}{\apj}, 682, 283

\bibitem[{{Goicoechea} {et~al.}(2004){Goicoechea},
  {Rodr{\'\i}guez-Fern{\'a}ndez}, \& {Cernicharo}}]{2004ApJ...600..214G}
{Goicoechea}, J.~R., {Rodr{\'\i}guez-Fern{\'a}ndez}, N.~J., \& {Cernicharo}, J.
  2004, \href{https://iopscience.iop.org/article/10.1086/379704/pdf}{\apj},
  600, 214

\bibitem[{{Hasegawa} \& {Herbst}(1993)}]{1993MNRAS.263..589H}
{Hasegawa}, T.~I. \& {Herbst}, E. 1993,
  \href{http://articles.adsabs.harvard.edu/pdf/1993MNRAS.263..589H}{\mnras},
  263, 589

\bibitem[{{Hassel} {et~al.}(2011){Hassel}, {Harada}, \&
  {Herbst}}]{2011ApJ...743..182H}
{Hassel}, G.~E., {Harada}, N., \& {Herbst}, E. 2011,
  \href{https://iopscience.iop.org/article/10.1088/0004-637X/743/2/182/pdf}{\apj},
  743, 182

\bibitem[{{Hassel} {et~al.}(2008){Hassel}, {Herbst}, \&
  {Garrod}}]{2008ApJ...681.1385H}
{Hassel}, G.~E., {Herbst}, E., \& {Garrod}, R.~T. 2008,
  \href{http://iopscience.iop.org/article/10.1086/588185/pdf}{\apj}, 681, 1385

\bibitem[{{Herbst} \& {van Dishoeck}(2009)}]{2009ARA&A..47..427H}
{Herbst}, E. \& {van Dishoeck}, E.~F. 2009,
  \href{https://www.annualreviews.org/doi/pdf/10.1146/annurev-astro-082708-101654}{\araa},
  47, 427

\bibitem[{{Hollis} {et~al.}(2004{\natexlab{a}}){Hollis}, {Jewell}, {Lovas}, \&
  {Remijan}}]{2004ApJ...613L..45H}
{Hollis}, J.~M., {Jewell}, P.~R., {Lovas}, F.~J., \& {Remijan}, A.
  2004{\natexlab{a}},
  \href{https://iopscience.iop.org/article/10.1086/424927/pdf}{\apjl}, 613, L45

\bibitem[{{Hollis} {et~al.}(2004{\natexlab{b}}){Hollis}, {Jewell}, {Lovas},
  {Remijan}, \& {M{\o}llendal}}]{2004ApJ...610L..21H}
{Hollis}, J.~M., {Jewell}, P.~R., {Lovas}, F.~J., {Remijan}, A., \&
  {M{\o}llendal}, H. 2004{\natexlab{b}},
  \href{https://iopscience.iop.org/article/10.1086/423200/pdf}{\apjl}, 610, L21

\bibitem[{{Hollis} {et~al.}(2006{\natexlab{a}}){Hollis}, {Lovas}, {Remijan},
  {Jewell}, {Ilyushin}, \& {Kleiner}}]{2006ApJ...643L..25H}
{Hollis}, J.~M., {Lovas}, F.~J., {Remijan}, A.~J., {et~al.} 2006{\natexlab{a}},
  \href{https://iopscience.iop.org/article/10.1086/505110/pdf}{\apjl}, 643, L25

\bibitem[{{Hollis} {et~al.}(2006{\natexlab{b}}){Hollis}, {Remijan}, {Jewell},
  \& {Lovas}}]{2006ApJ...642..933H}
{Hollis}, J.~M., {Remijan}, A.~J., {Jewell}, P.~R., \& {Lovas}, F.~J.
  2006{\natexlab{b}},
  \href{https://iopscience.iop.org/article/10.1086/501121/pdf}{\apj}, 642, 933

\bibitem[{{H{\"u}ttemeister} {et~al.}(1993){H{\"u}ttemeister}, {Wilson},
  {Henkel}, \& {Mauersberger}}]{1993A&A...276..445H}
{H{\"u}ttemeister}, S., {Wilson}, T.~L., {Henkel}, C., \& {Mauersberger}, R.
  1993,
  \href{http://articles.adsabs.harvard.edu/pdf/1993A%26A...276..445H}{\aap},
  276, 445

\bibitem[{{H{\"u}ttemeister} {et~al.}(1995){H{\"u}ttemeister}, {Wilson},
  {Mauersberger}, {Lemme}, {Dahmen}, \& {Henkel}}]{1995A&A...294..667H}
{H{\"u}ttemeister}, S., {Wilson}, T.~L., {Mauersberger}, R., {et~al.} 1995,
  \href{http://articles.adsabs.harvard.edu/pdf/1995A%26A...294..667H}{\aap},
  294, 667

\bibitem[{{Indriolo} {et~al.}(2015){Indriolo}, {Neufeld}, {Gerin}, {Schilke},
  {Benz}, {Winkel}, {Menten}, {Chambers}, {Black}, {Bruderer}, {Falgarone},
  {Godard}, {Goicoechea}, {Gupta}, {Lis}, {Ossenkopf}, {Persson},
  {Sonnentrucker}, {van der Tak}, {van Dishoeck}, {Wolfire}, \&
  {Wyrowski}}]{2015ApJ...800...40I}
{Indriolo}, N., {Neufeld}, D.~A., {Gerin}, M., {et~al.} 2015,
  \href{https://iopscience.iop.org/article/10.1088/0004-637X/800/1/40/pdf}{\apj},
  800, 40

\bibitem[{{Jaber} {et~al.}(2014){Jaber}, {Ceccarelli}, {Kahane}, \&
  {Caux}}]{2014ApJ...791...29J}
{Jaber}, A.~A., {Ceccarelli}, C., {Kahane}, C., \& {Caux}, E. 2014,
  \href{https://iopscience.iop.org/article/10.1088/0004-637X/791/1/29/pdf}{\apj},
  791, 29

\bibitem[{{Jim{\'e}nez-Serra} {et~al.}(2016){Jim{\'e}nez-Serra}, {Vasyunin},
  {Caselli}, {Marcelino}, {Billot}, {Viti}, {Testi}, {Vastel}, {Lefloch}, \&
  {Bachiller}}]{2016ApJ...830L...6J}
{Jim{\'e}nez-Serra}, I., {Vasyunin}, A.~I., {Caselli}, P., {et~al.} 2016,
  \href{https://iopscience.iop.org/article/10.3847/2041-8205/830/1/L6/pdf}{\apjl},
  830, L6

\bibitem[{{Jin} \& {Garrod}(2020)}]{2020ApJS..249...26J}
{Jin}, M. \& {Garrod}, R.~T. 2020,
  \href{https://iopscience.iop.org/article/10.3847/1538-4365/ab9ec8/pdf}{\apjs},
  249, 26

\bibitem[{{Jones} {et~al.}(2008){Jones}, {Burton}, {Cunningham}, {Menten},
  {Schilke}, {Belloche}, {Leurini}, {Ott}, \& {Walsh}}]{2008MNRAS.386..117J}
{Jones}, P.~A., {Burton}, M.~G., {Cunningham}, M.~R., {et~al.} 2008,
  \href{https://watermark.silverchair.com/mnras0386-0117.pdf?token=AQECAHi208BE49Ooan9kkhW_Ercy7Dm3ZL_9Cf3qfKAc485ysgAAAskwggLFBgkqhkiG9w0BBwagggK2MIICsgIBADCCAqsGCSqGSIb3DQEHATAeBglghkgBZQMEAS4wEQQM4CHzgjDQrHRleum0AgEQgIICfJbcvq4r1nymTLUtoD262CS5fSITCmu_YcayCrNMS3xaMaXPXgPtF-O1TgNw5v-eqLn9vi-YlqHwSAZzmR9I31-5ybTIs0o2eYlkV1FA_6t4WcHqLB1Dhr9RJPIn3GcZx5j7NY2w2dYzTnF-ASNAbL1N4nluORdapKAp2iK_VMhsYz-VCeBLFRWl-luJ18D_ZOpJKN3EI9MGaPMvAbn8thdL75Xwz2Fgdm048fPmLJNB1m-wvmuy7EbKvGz62p-gbZfkQj3IztyPio2GZqY0eiXFTLzlfcDf-LL-6Amnre2rqCV7-YWEsShjf-1wg09CobS7rym-3iwHoq-UpjgA4Oo--4bSK1NjlbrUXpNjvfQfdzZzdmMCIzcLfdmmQzNKH1Ydkj7g1DvZu2RUqHg7hhLu2dacrBGreDMOzvuSBT38jb11pfd9tjvaBbBAkR4fmMKU4zuE7bZCMwwyBhbzbNiHVHA5T_lB-f8CQaOBN6W1U_exccOulRivJpXikHu1nmXqJLPLl9UGXjhcjZ8zYPJp3Q-8gbJtTp95ljqmqA6OMwopEot8xNMOSvRdgj3GlP7glC0NbFPQxwf4h8jIeEVZo9l9SmFbQMEiwx16SoZhq7QO9uNEpSwqIA_gGrfly7zFkOVs8ipiK96ParnZe8WNc4n4eOb-h61mXOuA2GbCsYT0kM8ds7o3jB_aa00CxUHNroWuEZbfMrqPt5GflQlPtNyjyLDUEPTnLjX5AUhRJulphLo9AoCnv74XBRAV038Ryqm5XaRxKiiHR27Rrr4mH4olkZ4SMRDb_hjPK4Ok1yPxnGB_ljgdBL0POSpEpDCYhQs3C0VLky58qg}{\mnras},
  386, 117

\bibitem[{{Jones} {et~al.}(2011){Jones}, {Burton}, {Tothill}, \&
  {Cunningham}}]{2011MNRAS.411.2293J}
{Jones}, P.~A., {Burton}, M.~G., {Tothill}, N.~F.~H., \& {Cunningham}, M.~R.
  2011,
  \href{https://watermark.silverchair.com/mnras0411-2293.pdf?token=AQECAHi208BE49Ooan9kkhW_Ercy7Dm3ZL_9Cf3qfKAc485ysgAAAsowggLGBgkqhkiG9w0BBwagggK3MIICswIBADCCAqwGCSqGSIb3DQEHATAeBglghkgBZQMEAS4wEQQM4yV8qEhM26a6TDxeAgEQgIICfca51w_EPGv8B21RRIbZKey1bNBGbi28OP34Xtb-yBwg7KWKFmsBozh_88k2iLRZjiE_3hv6ILCDomlCVKrSxqeEi-_g2VuBFnvGIb_dYtQgrvlZBYrKlTGBpVYJTFtqlKZlfwntyCSZFMDwgvwi8-rGmV1bjR6kCNo9G_LTLc8eIxnsTEp_ru8fk3a0Pra9tPS5VaCellEckSdwi72qcJdRz5jbW9EmRWejRDIQqoA3gNOgx6JfltianRy0n9K2GOiGwSKxk_PnT83QX9rTxfB77RlsCcIA5tKdMouqXEfvWdWK3kflIb2xsAu1nZFlbrcQslhjyca3IwlJscOrAMDr7ivsSlaOF5VE41ASJ_Sey0SmCWjcRvIRn5Git0l_nqWohoPTmRg-oiTP1_Xeju7ThFMOXtfJ3e4Sxg8ma3i7co8Dj4rp9SepWkNcV_QFHGYs79Gk_JFSs8DljsEQLNt7KAXwj1P5vDsFGVyvvedNHN2Fai9PWkr3wH2jY2pz0oxhmKBXbMOqsBDiachXfUB_9kNvxWbqjtkUWJrcPVOaUF-MkJwvnacUz0-st0Eg7ih93huW1GIHYw9p0sv9gtk1bML9QeaJPiytNr4OPaSZIING3nwUOUdDXNrv0yOL372fwWcEAUMmRH_8NBMRyZupZCKGwc1kqf8dyWFvnS3yLS0q1L1DM3TRx2oj5RKbAY5F5ymG5DeJ0Z6e3xJ9RxXvXtOfI_Lg7IHaNDmuaF5T1xN3fZN47w9oXIG0IdT8hS_d0oTnIj4bp-3oV-pvApcmwvYTcouifGcibFMPzY6N6tzq7WptiyISzCD4AWtUq83KCkqHyzbrwJF03c8}{\mnras},
  411, 2293

\bibitem[{{J{\o}rgensen} {et~al.}(2020){J{\o}rgensen}, {Belloche}, \&
  {Garrod}}]{2020ARA&A..58..727J}
{J{\o}rgensen}, J.~K., {Belloche}, A., \& {Garrod}, R.~T. 2020,
  \href{https://www.annualreviews.org/doi/pdf/10.1146/annurev-astro-032620-021927}{\araa},
  58, 727

\bibitem[{{Le Petit} {et~al.}(2016){Le Petit}, {Ruaud}, {Bron}, {Godard},
  {Roueff}, {Languignon}, \& {Le Bourlot}}]{2016A&A...585A.105L}
{Le Petit}, F., {Ruaud}, M., {Bron}, E., {et~al.} 2016,
  \href{https://www.aanda.org/articles/aa/pdf/2016/01/aa26658-15.pdf}{\aap},
  585, A105

\bibitem[{{Li} {et~al.}(2020){Li}, {Wang}, {Qiao}, {Quan}, {Fang}, {Du}, {Li},
  {Shen}, {Li}, {Li}, {Shi}, {Zhang}, \& {Zhang}}]{2020MNRAS.492..556L}
{Li}, J., {Wang}, J., {Qiao}, H., {et~al.} 2020,
  \href{https://academic.oup.com/mnras/pdf-lookup/doi/10.1093/mnras/stz3337}{\mnras},
  492, 556

\bibitem[{{Liu} {et~al.}(2020){Liu}, {Chen}, \& {Du}}]{2020ApJ...899...92L}
{Liu}, C., {Chen}, X., \& {Du}, F. 2020,
  \href{https://iopscience.iop.org/article/10.3847/1538-4357/aba758/pdf}{\apj},
  899, 92

\bibitem[{{Loomis} {et~al.}(2013){Loomis}, {Zaleski}, {Steber}, {Neill},
  {Muckle}, {Harris}, {Hollis}, {Jewell}, {Lattanzi}, {Lovas}, {Martinez},
  {McCarthy}, {Remijan}, {Pate}, \& {Corby}}]{2013ApJ...765L...9L}
{Loomis}, R.~A., {Zaleski}, D.~P., {Steber}, A.~L., {et~al.} 2013,
  \href{https://iopscience.iop.org/article/10.1088/2041-8205/765/1/L9/pdf}{\apjl},
  765, L9

\bibitem[{{Lu} {et~al.}(2018){Lu}, {Chang}, \& {Aikawa}}]{2018ApJ...869..165L}
{Lu}, Y., {Chang}, Q., \& {Aikawa}, Y. 2018,
  \href{https://iopscience.iop.org/article/10.3847/1538-4357/aaeed8/pdf}{\apj},
  869, 165

\bibitem[{{Marrone} {et~al.}(2008){Marrone}, {Baganoff}, {Morris}, {Moran},
  {Ghez}, {Hornstein}, {Dowell}, {Mu{\~n}oz}, {Bautz}, {Ricker}, {Brandt},
  {Garmire}, {Lu}, {Matthews}, {Zhao}, {Rao}, \& {Bower}}]{2008ApJ...682..373M}
{Marrone}, D.~P., {Baganoff}, F.~K., {Morris}, M.~R., {et~al.} 2008,
  \href{https://iopscience.iop.org/article/10.1086/588806/pdf}{\apj}, 682, 373

\bibitem[{{McGuire}(2018)}]{2018ApJS..239...17M}
{McGuire}, B.~A. 2018,
  \href{https://iopscience.iop.org/article/10.3847/1538-4365/aae5d2/pdf}{\apjs},
  239, 17

\bibitem[{{Nobukawa} {et~al.}(2011){Nobukawa}, {Ryu}, {Tsuru}, \&
  {Koyama}}]{2011ApJ...739L..52N}
{Nobukawa}, M., {Ryu}, S.~G., {Tsuru}, T.~G., \& {Koyama}, K. 2011,
  \href{https://iopscience.iop.org/article/10.1088/2041-8205/739/2/L52/pdf}{\apjl},
  739, L52

\bibitem[{{{\"O}berg} {et~al.}(2010){{\"O}berg}, {Bottinelli}, {J{\o}rgensen},
  \& {van Dishoeck}}]{2010ApJ...716..825O}
{{\"O}berg}, K.~I., {Bottinelli}, S., {J{\o}rgensen}, J.~K., \& {van Dishoeck},
  E.~F. 2010,
  \href{https://iopscience.iop.org/article/10.1088/0004-637X/716/1/825/pdf}{\apj},
  716, 825

\bibitem[{{{\"O}berg} {et~al.}(2007){{\"O}berg}, {Fuchs}, {Awad}, {Fraser},
  {Schlemmer}, {van Dishoeck}, \& {Linnartz}}]{2007ApJ...662L..23O}
{{\"O}berg}, K.~I., {Fuchs}, G.~W., {Awad}, Z., {et~al.} 2007,
  \href{http://iopscience.iop.org/article/10.1086/519281/pdf}{\apjl}, 662, L23

\bibitem[{{{\"O}berg} {et~al.}(2009{\natexlab{a}}){{\"O}berg}, {Linnartz},
  {Visser}, \& {van Dishoeck}}]{2009ApJ...693.1209O}
{{\"O}berg}, K.~I., {Linnartz}, H., {Visser}, R., \& {van Dishoeck}, E.~F.
  2009{\natexlab{a}},
  \href{https://iopscience.iop.org/article/10.1088/0004-637X/693/2/1209/pdf}{\apj},
  693, 1209

\bibitem[{{{\"O}berg} {et~al.}(2009{\natexlab{b}}){{\"O}berg}, {van Dishoeck},
  \& {Linnartz}}]{2009A&A...496..281O}
{{\"O}berg}, K.~I., {van Dishoeck}, E.~F., \& {Linnartz}, H.
  2009{\natexlab{b}},
  \href{https://www.aanda.org/articles/aa/pdf/2009/10/aa10207-08.pdf}{\aap},
  496, 281

\bibitem[{{Potapov} {et~al.}(2020){Potapov}, {J{\"a}ger}, \&
  {Henning}}]{2020PhRvL.124v1103P}
{Potapov}, A., {J{\"a}ger}, C., \& {Henning}, T. 2020,
  \href{https://journals.aps.org/prl/abstract/10.1103/PhysRevLett.124.221103}{\prl},
  124, 221103

\bibitem[{{Rea} {et~al.}(2013){Rea}, {Esposito}, {Pons}, {Turolla}, {Torres},
  {Israel}, {Possenti}, {Burgay}, {Vigan{\`o}}, {Papitto}, {Perna}, {Stella},
  {Ponti}, {Baganoff}, {Haggard}, {Camero-Arranz}, {Zane}, {Minter},
  {Mereghetti}, {Tiengo}, {Sch{\"o}del}, {Feroci}, {Mignani}, \&
  {G{\"o}tz}}]{2013ApJ...775L..34R}
{Rea}, N., {Esposito}, P., {Pons}, J.~A., {et~al.} 2013,
  \href{https://iopscience.iop.org/article/10.1088/2041-8205/775/2/L34/pdf}{\apjl},
  775, L34

\bibitem[{{Reid} {et~al.}(2014){Reid}, {Menten}, {Brunthaler}, {Zheng}, {Dame},
  {Xu}, {Wu}, {Zhang}, {Sanna}, {Sato}, {Hachisuka}, {Choi}, {Immer},
  {Moscadelli}, {Rygl}, \& {Bartkiewicz}}]{2014ApJ...783..130R}
{Reid}, M.~J., {Menten}, K.~M., {Brunthaler}, A., {et~al.} 2014,
  \href{https://iopscience.iop.org/article/10.1088/0004-637X/783/2/130/pdf}{\apj},
  783, 130

\bibitem[{{Remijan} {et~al.}(2008){Remijan}, {Hollis}, {Lovas}, {Stork},
  {Jewell}, \& {Meier}}]{2008ApJ...675L..85R}
{Remijan}, A.~J., {Hollis}, J.~M., {Lovas}, F.~J., {et~al.} 2008,
  \href{https://iopscience.iop.org/article/10.1086/533529/pdf}{\apjl}, 675, L85

\bibitem[{{Requena-Torres} {et~al.}(2008){Requena-Torres},
  {Mart{\'\i}n-Pintado}, {Mart{\'\i}n}, \& {Morris}}]{2008ApJ...672..352R}
{Requena-Torres}, M.~A., {Mart{\'\i}n-Pintado}, J., {Mart{\'\i}n}, S., \&
  {Morris}, M.~R. 2008,
  \href{https://iopscience.iop.org/article/10.1086/523627/pdf}{\apj}, 672, 352

\bibitem[{{Requena-Torres} {et~al.}(2006){Requena-Torres},
  {Mart{\'\i}n-Pintado}, {Rodr{\'\i}guez-Franco}, {Mart{\'\i}n},
  {Rodr{\'\i}guez-Fern{\'a}ndez}, \& {de Vicente}}]{2006A&A...455..971R}
{Requena-Torres}, M.~A., {Mart{\'\i}n-Pintado}, J., {Rodr{\'\i}guez-Franco},
  A., {et~al.} 2006,
  \href{https://www.aanda.org/articles/aa/pdf/2006/33/aa5190-06.pdf}{\aap},
  455, 971

\bibitem[{{Roberge} {et~al.}(1991){Roberge}, {Jones}, {Lepp}, \&
  {Dalgarno}}]{1991ApJS...77..287R}
{Roberge}, W.~G., {Jones}, D., {Lepp}, S., \& {Dalgarno}, A. 1991,
  \href{http://articles.adsabs.harvard.edu/pdf/1991ApJS...77..287R}{\apjs}, 77,
  287

\bibitem[{{Rodr{\'\i}guez-Fern{\'a}ndez}
  {et~al.}(2004){Rodr{\'\i}guez-Fern{\'a}ndez}, {Mart{\'\i}n-Pintado},
  {Fuente}, \& {Wilson}}]{2004A&A...427..217R}
{Rodr{\'\i}guez-Fern{\'a}ndez}, N.~J., {Mart{\'\i}n-Pintado}, J., {Fuente}, A.,
  \& {Wilson}, T.~L. 2004,
  \href{https://www.aanda.org/articles/aa/pdf/2004/43/aa1370.pdf}{\aap}, 427,
  217

\bibitem[{{Ruaud} {et~al.}(2015){Ruaud}, {Loison}, {Hickson}, {Gratier},
  {Hersant}, \& {Wakelam}}]{2015MNRAS.447.4004R}
{Ruaud}, M., {Loison}, J.~C., {Hickson}, K.~M., {et~al.} 2015,
  \href{https://watermark.silverchair.com/stu2709.pdf?token=AQECAHi208BE49Ooan9kkhW_Ercy7Dm3ZL_9Cf3qfKAc485ysgAAArgwggK0BgkqhkiG9w0BBwagggKlMIICoQIBADCCApoGCSqGSIb3DQEHATAeBglghkgBZQMEAS4wEQQMVlWFa1rbPZ_-pkxTAgEQgIICa1lRwUvSSkxkt15J2ra5Zxb1B346X8odVFKddC-_qFvMw3IRqxm9lRO2eFcM8AplGfSIJirhliZ2SglISSpMwPuYvu2XQGKO8_VpcYli6ioaJJ14_naoVqNUly7vId63EPhdkyZ5wzxug2ZSL-xaNWku4YBJNnBEtfMdNSviGmpppWzx50jcAmofR6ZYjaBLSgpvE3MaO2XgeblxS60QvF_C79VsZ0kwEd19Ci4Vt7-Mt0udGhqem40hXyI6Wp6wWgvicMI1fjbVvO4ck0xQJXHUxQNzDlmqXiiwHQioAF_IFaFEsTy26lw5cCgJmUXfkTX7ImjYSdL7OTQYGLylQedxHIfcu35XpTLX0YYjccaam5dN5Hfe_XB3MikUh3xSrri0l3DKfsgvA1vXqwkZPm-I4eOHEuBlWzBTPkmEv0ooriAdh39MqALXv9Fm3ZRNJUDZpAg8wxcfK1mULDxH7UTftcKgUYa72elJ1IzVdzY6iRTaADMxGvnCIuNsuz9bDJ_ARPzyCKmMETmQwnGUZJhhDkhVaIMrh-i2FP1BAgLL0ffoTTHWXJhWWmqufcef4bXFaQqZoL2dhMXWOht2--zyTKUf86E18Mwr9myBJOTReC1tpMDYL3OArldJrbTnMg-cQkXk9jsFenZqcDnNXpBvoD-secR-uVBVmzzAsN4Xsf0QN8onfy1vsumvGjsxB7my6Ur1CKvdECL50-Vt05tooRJcksN5SV5xAoFZtk0j1hGxhGdSkRsfTXHLq8WJnL2_d36-Yp5iyI2BifUJi5-5th5brUp-X3UdfKvx1qbuwTStKDRplqrwfvw}{\mnras},
  447, 4004

\bibitem[{{Ryu} {et~al.}(2013){Ryu}, {Nobukawa}, {Nakashima}, {Tsuru},
  {Koyama}, \& {Uchiyama}}]{2013PASJ...65...33R}
{Ryu}, S.~G., {Nobukawa}, M., {Nakashima}, S., {et~al.} 2013,
  \href{https://watermark.silverchair.com/pasj65-0033.pdf?token=AQECAHi208BE49Ooan9kkhW_Ercy7Dm3ZL_9Cf3qfKAc485ysgAAArcwggKzBgkqhkiG9w0BBwagggKkMIICoAIBADCCApkGCSqGSIb3DQEHATAeBglghkgBZQMEAS4wEQQM1yYLId0rHpJE8kOlAgEQgIICagH3KkPRcVWgWa2Jk1gQToqJHBDFMmy_v5d3xPvEcRUU7cwg9tfl3t_6rcQqiTPmW6NGuD8EzEvTwbiqBj4_9GCyUT4D-idCQAJkz3ynFzcA87d4fTKXnzMfJp0rIjoio1-25ncrfGK-CF9tC27DmNT3Yg7twg5mlbWiu2fBtHG3i-rlfI-80mT0sWle7FgdBatJLNl-t6_f6OhUlofRTbo90F-KcAT51KANlj58R4Rh0HA1ftimXZscKvY7yDJN5znqDAfrkDeSBP4mxsjCJx23dVFo810RVlU0uj_keoNvgln6v5sqvqNulNnMjaybJfs-f1zbPx5dz6g1afCzdX1fbXD6e4wcgC9m1JOMODvS7zg1KaocJSdSBc5azy290FLwtQxK2iB-Wkb3tnRykUM6Pb9DZBSfIE5CF63qfVyVzLIAApw362AUliP0_xbq5T_7JUlXukINGmLKZr5YCXEDQnOjAW9V1lRNM8Avee8aKBEa-x_yoLz8TQzPB5YWRA29REglmZztSeUt6Vo32B2tFvRr__UkXyhfD3o3qdlVcdXS8DshoCxmtQq7KajZHhphegU-gWEE6yXDBY1JNB0PnAaC7ZQWzOOgI6BW11-rBr6vYKyiDLO_13UjYO8eL68f-JL1aeL5OVvOLxH27bWjRzr8-nLDXNdbmba_IU_BTiKnHQ9_iO4C-DDMgwN3-C_ydjvcaf12_pgENAsOBxZ_iJxIwp2nxfFfTwJEvi8GlFCocqgZR-4kPNfaWwuAqlCxxx2HRU6f5t7HTtzkiiL5hcqj3hPTpKtmGDzIcoLBgqZDwwQqf5bVXQ}{\pasj},
  65, 33

\bibitem[{{Schmiedeke} {et~al.}(2016){Schmiedeke}, {Schilke}, {M{\"o}ller},
  {S{\'a}nchez-Monge}, {Bergin}, {Comito}, {Csengeri}, {Lis}, {Molinari},
  {Qin}, \& {Rolffs}}]{2016A&A...588A.143S}
{Schmiedeke}, A., {Schilke}, P., {M{\"o}ller}, T., {et~al.} 2016,
  \href{https://www.aanda.org/articles/aa/pdf/2016/04/aa27311-15.pdf}{\aap},
  588, A143

\bibitem[{{Shen} {et~al.}(2004){Shen}, {Greenberg}, {Schutte}, \& {van
  Dishoeck}}]{2004A&A...415..203S}
{Shen}, C.~J., {Greenberg}, J.~M., {Schutte}, W.~A., \& {van Dishoeck}, E.~F.
  2004, \href{https://www.aanda.org/articles/aa/pdf/2004/07/aa3457.pdf}{\aap},
  415, 203

\bibitem[{{Shingledecker} {et~al.}(2018){Shingledecker}, {Tennis}, {Le Gal}, \&
  {Herbst}}]{2018ApJ...861...20S}
{Shingledecker}, C.~N., {Tennis}, J., {Le Gal}, R., \& {Herbst}, E. 2018,
  \href{https://iopscience.iop.org/article/10.3847/1538-4357/aac5ee/pdf}{\apj},
  861, 20

\bibitem[{{Spitzer} \& {Tomasko}(1968)}]{1968ApJ...152..971S}
{Spitzer}, Lyman, J. \& {Tomasko}, M.~G. 1968,
  \href{http://articles.adsabs.harvard.edu/pdf/1968ApJ...152..971S}{\apj}, 152,
  971

\bibitem[{{Terrier} {et~al.}(2010){Terrier}, {Ponti}, {B{\'e}langer},
  {Decourchelle}, {Tatischeff}, {Goldwurm}, {Trap}, {Morris}, \&
  {Warwick}}]{2010ApJ...719..143T}
{Terrier}, R., {Ponti}, G., {B{\'e}langer}, G., {et~al.} 2010,
  \href{https://iopscience.iop.org/article/10.1088/0004-637X/719/1/143/pdf}{\apj},
  719, 143

\bibitem[{{Vastel} {et~al.}(2014){Vastel}, {Ceccarelli}, {Lefloch}, \&
  {Bachiller}}]{2014ApJ...795L...2V}
{Vastel}, C., {Ceccarelli}, C., {Lefloch}, B., \& {Bachiller}, R. 2014,
  \href{https://iopscience.iop.org/article/10.1088/2041-8205/795/1/L2/pdf}{\apjl},
  795, L2

\bibitem[{{Vasyunin} \& {Herbst}(2013{\natexlab{a}})}]{2013ApJ...762...86V}
{Vasyunin}, A.~I. \& {Herbst}, E. 2013{\natexlab{a}},
  \href{http://iopscience.iop.org/article/10.1088/0004-637X/762/2/86/pdf}{\apj},
  762, 86

\bibitem[{{Vasyunin} \& {Herbst}(2013{\natexlab{b}})}]{2013ApJ...769...34V}
{Vasyunin}, A.~I. \& {Herbst}, E. 2013{\natexlab{b}},
  \href{https://iopscience.iop.org/article/10.1088/0004-637X/769/1/34/pdf}{\apj},
  769, 34

\bibitem[{{Vasyunin} {et~al.}(2009){Vasyunin}, {Semenov}, {Wiebe}, \&
  {Henning}}]{2009ApJ...691.1459V}
{Vasyunin}, A.~I., {Semenov}, D.~A., {Wiebe}, D.~S., \& {Henning}, T. 2009,
  \href{http://iopscience.iop.org/article/10.1088/0004-637X/691/2/1459/pdf}{\apj},
  691, 1459

\bibitem[{{Wang} {et~al.}(2019){Wang}, {Chang}, \&
  {Wang}}]{2019A&A...622A.185W}
{Wang}, Y., {Chang}, Q., \& {Wang}, H. 2019,
  \href{https://www.aanda.org/articles/aa/pdf/2019/02/aa34276-18.pdf}{\aap},
  622, A185

\bibitem[{{Willis} {et~al.}(2020){Willis}, {Garrod}, {Belloche}, {M{\"u}ller},
  {Barger}, {Bonfand}, \& {Menten}}]{2020A&A...636A..29W}
{Willis}, E.~R., {Garrod}, R.~T., {Belloche}, A., {et~al.} 2020,
  \href{https://www.aanda.org/articles/aa/pdf/2020/04/aa36489-19.pdf}{\aap},
  636, A29

\bibitem[{{Zaleski} {et~al.}(2013){Zaleski}, {Seifert}, {Steber}, {Muckle},
  {Loomis}, {Corby}, {Martinez}, {Crabtree}, {Jewell}, {Hollis}, {Lovas},
  {Vasquez}, {Nyiramahirwe}, {Sciortino}, {Johnson}, {McCarthy}, {Remijan}, \&
  {Pate}}]{2013ApJ...765L..10Z}
{Zaleski}, D.~P., {Seifert}, N.~A., {Steber}, A.~L., {et~al.} 2013,
  \href{https://iopscience.iop.org/article/10.1088/2041-8205/765/1/L10/pdf}{\apjl},
  765, L10

\end{thebibliography}

                
\end{document}